\documentstyle[11pt,epsf]{article}

\begin{document}

\title{{\bf SU(2) and SU(1,1) Algebra Eigenstates:\\
A Unified Analytic Approach to Coherent \\
and Intelligent States}}
\author{{\bf Constantin Brif\ }\thanks{
E-mail: costya@physics.technion.ac.il. 
URL address: http://www.technion.ac.il/\~{}brif.} \\
{\em Department of Physics, 
Technion -- Israel Institute of Technology, Haifa 32000, Israel}}
\date{}
\maketitle

%\setlength{\baselineskip}{0.73cm}
%\newpage

\begin{abstract}
We introduce the concept of algebra eigenstates which are defined
for an arbitrary Lie group as eigenstates of elements of the 
corresponding complex Lie algebra. We show that this concept unifies
different definitions of coherent states associated with a dynamical 
symmetry group. On the one hand, algebra eigenstates include
different sets of Perelomov's generalized coherent states. On the
other hand, intelligent states (which are squeezed states for a 
system of general symmetry) also form a subset of algebra 
eigenstates. We develop the general formalism and apply it to the 
SU(2) and SU(1,1) simple Lie groups. Complete solutions to the 
general eigenvalue problem are found in the both cases, by a method
that employs analytic representations of the algebra eigenstates.
This analytic method also enables us to obtain exact closed
expressions for quantum statistical properties of an arbitrary
algebra eigenstate. Important special cases such as standard 
coherent states and intelligent states are examined and relations 
between them are studied by using their analytic representations. 
\end{abstract}

%\vspace{0.3cm}
%{\bf Keywords}: Coherent states, Intelligent states, 
%Analytic representations
%
%\newpage

\section{\uppercase{Introduction}}
\label{sec:Int}
\setcounter{equation}{0}

Coherent states (CS) associated with various Lie groups have been 
successfully used in the last decades in many problems of quantum 
physics (Klauder and Skagerstam, 1985; Perelomov, 1986; Zhang 
{\em et al.}, 1990). There are three different group-theoretic 
approaches to CS (Zhang {\em et al.}, 1990). These approaches
follow three possible definitions of the familiar Glauber CS 
$|\alpha\rangle$ of a harmonic oscillator (Glauber, 1963). Perelomov 
(1972, 1977), Gilmore (1972, 1974) and Rasetti (1975) have 
developed the formalism in which CS are generated by the action of 
group elements on a reference state of a group representation Hilbert 
space. In the second approach one deals with eigenstates of a 
lowering group generator (Barut and Girardello, 1971). The third 
definition of CS is related to the optimization of uncertainty 
relations for Hermitian generators of a group (Schr\"{o}dinger, 1926). 
States that minimize uncertainty relations are called intelligent 
states (IS) (Aragone {\em et al.}, 1974, 1976; Ruschin and Ben-Aryeh, 
1976; Vanden-Bergh and DeMeyer, 1978) or minimum-uncertainty states. 
Different definitions coincide only in the special case of the 
Heisenberg-Weyl group (Weyl, 1950) that is the dynamical symmetry group 
of a quantized harmonic oscillator; then one obtains the Glauber CS 
$|\alpha\rangle$ (Glauber, 1963). For more complicated groups, e.g., 
for SU(2) and SU(1,1), the different approaches lead to distinct states. 

Relations between various sets of coherent and intelligent states have 
been studied in a number of recent works (Wodkiewicz and Eberly, 1985;
Trifonov, 1994; Brif and Ben-Aryeh, 1994a).
In the present paper we continue and extend this study by developing 
a group-theoretic formalism that provides a unified description of 
different types of coherent and intelligent states. We introduce the 
concept of algebra eigenstates (AES) which are defined for an 
arbitrary Lie group as eigenstates of elements of the corresponding 
complex Lie algebra. This general approach incorporates in a simple 
way the three different definitions mentioned above.
Different sets of Perelomov's CS associated with a Lie group can be
equivalently defined as the AES for this group. 
The IS for Hermitian generators of a group also form a subset
of the AES. The algebra-eigenstate formalism enables us to use 
powerful analytic methods for treating different types of states in 
a unified way. 
Recently, we have also considered (Brif, 1996) the two-photon AES 
which provide a unified analytic approach to single-mode squeezing.
Similar ideas have been also discussed recently by Puri and Agarwal
(1996) and by Trifonov (1996a,b).

In this work we apply the general formalism to the SU(2) and SU(1,1)
simple Lie groups. We use analytic representations based on the 
standard sets of Perelomov's CS for these groups. In the SU(1,1) case
an alternative analytic representation based on the Barut-Girardello
states (Barut and Girardello, 1971) is also used. The eigenvalue 
equation that determines the AES is converted by means of an analytic 
representation into a linear homogeneous differential equation.
Solving this equation, we obtain the complete solution of the general
eigenvalue problem. Then the theory of analytic functions is applied 
for studying quantum statistical properties of the AES and relations 
between their subsets.

\section{\uppercase{The general theory of the algebra eigenstates}}
\label{sec:Gen}
\setcounter{equation}{0}

\subsection{Definitions}

Let $G$ be an arbitrary Lie group and $T$ its unitary irreducible 
representation acting on the Hilbert space ${\cal H}$. 
Let $Y$ be the complex Lie algebra of the group $G$ (in what
follows we will call algebra the complex extension of the real 
algebra, i.e., the set of all linear combinations of elements of the 
real algebra with complex coefficients). If we choose a basis 
$\{ y_{1},y_{2},\ldots,y_{p} \}$ for the $p$-dimensional Lie algebra 
$Y$, then an element of the complex algebra can be written as the 
Euclidean scalar product in the $p$-dimensional vector space,
\begin{equation}
(\vec{\beta}\cdot\vec{y}) = \beta_{1} y_{1} + \beta_{2} y_{2} + 
\cdots + \beta_{p} y_{p} \ ,   
\end{equation}
where $\beta_{1},\beta_{2},\ldots,\beta_{p}$ are arbitrary complex 
coefficients. Then, the AES are defined by the eigenvalue equation:
\begin{equation}
(\vec{\beta}\cdot\vec{y}) |\Psi(\lambda,\vec{\beta})\rangle = 
\lambda |\Psi(\lambda,\vec{\beta})\rangle \ , \;\;\;\;\;
|\Psi(\lambda,\vec{\beta})\rangle \in {\cal H} \ .   \label{aes:def}
\end{equation}
Admissible values of $\vec{\beta}$ and $\lambda$ depend on the
structure of the group and will be determined
for all particular situations that will be considered in the text. 
A special case of the eigenvalue equation (\ref{aes:def}) is the 
time-independent Schr\"{o}dinger equation for Hamiltonians which are 
linear combinations of group generators. However, we will see that 
apart from this special case the eigenvalue equation (\ref{aes:def}) 
contains other important cases and has a fundamental meaning in the 
theory of coherent and intelligent states.

\subsection{Generalized coherent states}

Now we turn back to the Perelomov definition of the generalized CS.
By choosing a fixed normalized reference state 
$|\Psi_{0}\rangle \in {\cal H}$, one can define the system of states 
$\{ |\Psi_{g}\rangle \}$, 
\begin{equation}
|\Psi_{g}\rangle = T(g) |\Psi_{0}\rangle \ ,
\mbox{\hspace{1cm}} g \in G \ , 
\end{equation}
which is called the coherent-state system. 
The isotropy (or maximum-stability) subgroup $H \subset G$ consists 
of all the group elements $h$ that leave the reference state 
invariant up to a phase factor,
\begin{equation}
T(h) |\Psi_{0}\rangle = e^{i\phi(h)} |\Psi_{0}\rangle \ , 
\mbox{\hspace{1cm}} | e^{i\phi(h)} | =1 \ ,
\mbox{\hspace{0.3cm}} h \in H \ .		\label{2.1}
\end{equation}
For every element $g \in G$, there is a unique decomposition of $g$ 
into a product of two group elements, one in $H$ and the other in the 
quotient (or coset) space $G/H$,
\begin{equation}
g = \Omega h \ , \mbox{\hspace{1cm}} g \in G, \;\; h \in H, \;\; 
\Omega \in G/H \ .	\label{2.2}	
\end{equation}
It is clear that group elements $g$ and $g'$ with different $h$ and 
$h'$ but with the same $\Omega$ produce the coherent states which 
differ only by a phase factor: 
$|\Psi_{g}\rangle = e^{i\delta} |\Psi_{g'}\rangle$, where 
$\delta =\phi(h) -\phi(h')$. 
Therefore a coherent state $|\Psi_{\Omega}\rangle$ is determined by 
a point $\Omega = \Omega(g)$ in the quotient space $G/H$. 

One can see from this group-theoretic procedure for the generation 
of the CS, that the choice of the reference state 
$|\Psi_{0}\rangle$ firmly determines the structure of the 
coherent-state set. Different choices of the reference state give
different sets of the CS.
An important class of coherent-state sets 
corresponds to the quotient spaces $G/H$ which are homogeneous
K\"{a}hlerian manifolds. Then $G/H$ can be considered as the phase 
space of a classical dynamical system, and the mapping 
$\Omega \rightarrow |\Psi_{\Omega}\rangle\langle\Psi_{\Omega}|$ is 
the quantization for this system (Berezin, 1975). The usually
used sets of the CS (the standard sets, as we refer to them)
correspond to the cases when an extreme state of the representation
Hilbert space (e.g., the vacuum state of an oscillator or the lowest
spin state) is chosen as the reference state (Zhang {\em et al.}, 
1990). In general, this choice of the reference state leads to the 
sets consisting of states with properties closest to those of 
classical states (Perelomov, 1986).

The isotropy subalgebra $X$ is defined as the set of 
elements $\{ x \}$, $x \in Y$, such that
\begin{equation}
x |\Psi_{0}\rangle = \lambda |\Psi_{0}\rangle \ ,
\label{2.3}
\end{equation}
where $\lambda$ is a complex eigenvalue. By acting with $T(g)$ on both 
sides of equation (\ref{2.3}), we obtain
\begin{equation}
T(g) x T^{-1}(g) T(g) |\Psi_{0}\rangle = \lambda T(g) 
|\Psi_{0}\rangle  \ .
\end{equation}
This leads to the eigenvalue equation
\begin{equation}
y |\Psi_{g}\rangle = \lambda |\Psi_{g}\rangle \ , \label{2.4}
\end{equation}
where $|\Psi_{g}\rangle = T(g) |\Psi_{0}\rangle$ is a coherent state, 
and $y = T(g) x T^{-1}(g)$ is an element of the algebra $Y$. We see 
that, for nontrivial $X$, Perelomov's generalized CS 
$|\Psi_{g}\rangle$ can be defined as the AES, and a specific set of 
the CS is obtained for the appropriate choice of the parameters 
$\beta$'s. More precisely, let a state 
$|\Psi(\lambda,\vec{\beta})\rangle$ belong to a specific set of the 
CS corresponding to the reference state $|\Psi_{0}\rangle$ that 
satisfies equation (\ref{2.3}). Then the parameters $\beta$'s 
must satisfy the condition 
$(\vec{\beta}\cdot\vec{y}) = T(g) x T^{-1}(g)$, 
$\forall g \in G$.
Note that the definition of the AES does not depend explicitly on 
the choice of the reference state $|\Psi_{0}\rangle$. Hence it is 
possible to treat the CS defined as the AES in a quite general way, 
regardless of the set to which they belong.

\subsection{Analytic representations}

An important property of the generalized CS is the identity 
resolution:
\begin{equation}
\int d\mu(\Omega) |\Psi_{\Omega}\rangle\langle\Psi_{\Omega}| = I \ ,   
\label{2.7}
\end{equation}
where $I$ is the identity operator in the Hilbert space ${\cal H}$, 
and $d\mu(\Omega)$ is the invariant measure in the homogeneous 
quotient space $G/H$. Then any state $|\Psi\rangle \in {\cal H}$ can 
be expanded in the coherent-state basis $|\Psi_{\Omega}\rangle$,
\begin{equation}
|\Psi\rangle = \int d\mu(\Omega) f(\Omega) |\Psi_{\Omega}\rangle \ ,   
\label{2.8}
\end{equation}
where $f(\Omega) = \langle\Psi_{\Omega}|\Psi\rangle$, and
\begin{equation}
\langle\Psi|\Psi\rangle = \int d\mu(\Omega) |f(\Omega)|^{2} < \infty \ .  
\label{2.9}
\end{equation}

Now, let us represent all the AES in the standard coherent-state 
basis. In what follows we will consider only the simplest cases 
in which the quotient space $G/H$ corresponding to the standard set 
is a homogeneous K\"{a}hlerian manifold that can be 
parametrized by a single complex number $\zeta$, so we write the 
standard generalized CS $|\Psi_{\Omega}\rangle$ in the form 
$|\zeta\rangle$. Then equation (\ref{2.8}) reads for the AES: 
\begin{equation}
|\Psi(\lambda,\vec{\beta})\rangle = \int d\mu(\zeta) 
f(\lambda,\vec{\beta};\zeta^{\ast}) |\zeta\rangle  \ .  \label{2.10}  
\end{equation}
The function $f(\lambda,\vec{\beta};\zeta) = 
\langle\zeta^{\ast}|\Psi(\lambda,\vec{\beta})\rangle$ can be 
factorized as $f(\lambda,\vec{\beta};\zeta) = 
{\cal R}(\zeta) \Lambda(\lambda,\vec{\beta};\zeta)$. Here 
${\cal R}(\zeta)$ is a normalization factor such that 
$\Lambda(\lambda,\vec{\beta};\zeta)$ is an analytic function of 
$\zeta$ defined on the whole complex plane or on part of it. Such 
analytic representations are well studied (Fock, 1928; Bargmann, 1961; 
Segal, 1962) for the standard coherent-state bases of the simplest Lie 
groups (Perelomov, 1986). In these simplest cases the 
elements of the Lie algebra act in the Hilbert space of 
analytic functions as linear differential operators. Then the 
eigenvalue equation (\ref{aes:def}) is converted into a linear 
homogeneous differential equation. Solving it, we obtain the 
analytic functions $\Lambda(\lambda,\vec{\beta};\zeta)$ representing 
the AES $|\Psi(\lambda,\vec{\beta})\rangle$ in the standard 
coherent-state basis $|\zeta\rangle$. 
The requirement of the analyticity provides us with the domain of 
admissible values of $\lambda$ and $\vec{\beta}$.
The knowledge of the function $\Lambda(\lambda,\vec{\beta};\zeta)$ 
enables us to calculate properties of the corresponding state. 
In this paper we will use these analytic representations for finding 
the expansion of the AES in the orthonormal basis of the representation
Hilbert space, including the explicit calculation of the normalization 
factor. We also will demonstrate how this analytic method can be used
for obtaining exact analytic expressions for some expectation values
over the AES.

\subsection{Intelligent states}

The standard set of Perelomov's CS is a particular case of the wide 
system of the AES. Other particular cases of the AES are the sets of 
the ordinary and generalized IS. Any two quantum observables 
(Hermitian operators in the Hilbert space) $A$ and $B$ obey the 
Schr\"{o}dinger-Robertson uncertainty relation 
\begin{equation}
(\Delta A)^{2} (\Delta B)^{2} \geq \frac{1}{4} 
( |\langle[A,B]\rangle|^{2} + 4 \sigma_{AB}^{2} ) \ ,   \label{2.16}
\end{equation}
where the variance of $A$ is $(\Delta A)^{2} = \langle A^{2} \rangle
- \langle A \rangle^{2}$, $(\Delta B)^{2}$ is defined similarly, the 
covariance of $A$ and $B$ is $\sigma_{AB} = \frac{1}{2} 
\langle AB+BA \rangle - \langle A \rangle \langle B \rangle$, and the 
expectation values are taken over an arbitrary state in the Hilbert 
space. When the covariance of $A$ and $B$ vanishes, 
$\sigma_{AB} = 0$, the Schr\"{o}dinger-Robertson uncertainty relation 
is reduced to the Heisenberg uncertainty relation,
\begin{equation}
(\Delta A)^{2} (\Delta B)^{2} \geq \frac{1}{4} 
|\langle[A,B]\rangle|^{2} \ .   \label{2.17}
\end{equation}
The ordinary IS (Aragone {\em et al.}, 1974, 1976) provide an equality 
in the Heisenberg uncertainty relation (\ref{2.17}), while the 
generalized IS (Trifonov, 1994) do so in the Schr\"{o}dinger-Robertson
uncertainty relation (\ref{2.16}). 
It is clear that the ordinary IS form a subset of the generalized IS. 
The generalized IS for operators $A$ and $B$ are determined 
from the eigenvalue equation (Trifonov, 1994; Puri, 1994)
\begin{equation}
(\eta A + i B) |\lambda,\eta\rangle = \lambda |\lambda,\eta\rangle \ , 
\label{2.18}
\end{equation}
where the parameter $\eta$ is an arbitrary complex number, and 
$\lambda$ is a complex eigenvalue. For the particular case of real 
$\eta$, the eigenvalue equation (\ref{2.18}) determines the ordinary 
IS for operators $A$ and $B$. Then the equation can be written in 
the form (Jackiw, 1968)
\begin{equation}
(A + i\gamma B) |\lambda,\gamma\rangle = 
\lambda |\lambda,\gamma\rangle \ ,   \label{2.19}
\end{equation}
where $\gamma$ is a real parameter. By comparing equations 
(\ref{2.18}) and (\ref{2.19}) with equation (\ref{aes:def}), we see 
that the IS for any two Hermitian group generators form a subset of 
the AES of the group. Eigenstates of a lowering group generator 
(Barut and Girardello, 1971; Dodonov {\em et al.}, 1974; Hillery, 
1987, 1989; Agarwal, 1988; Bu\v{z}ek, 1990; Brif and Ben-Aryeh, 1994b;
Brif, 1995), being a special case of the IS, are a simple example of 
the AES. 

The generalized IS for position and momentum of a harmonic oscillator 
coincide (Trifonov, 1994) with the canonical squeezed states (Stoler, 
1970, 1971; Yuen, 1976), and they also are referred to as the 
correlated coherent states (Dodonov {\em et al.}, 1980).
The concept of squeezing is naturally related also to the IS associated 
with more complicated Lie groups (Wodkiewicz and Eberly, 1985;
Hillery, 1987, 1989; Nieto and Truax, 1993; Trifonov, 1994).
At the last years there is a great interest in the IS (Wodkiewicz and 
Eberly, 1985; Agarwal and Puri, 1990; Bergou {\em et al.}, 1991;
Hillery and Mlodinow, 1993; Nieto and Truax, 1993; Trifonov, 1994; 
Brif and Ben-Aryeh, 1994a; Yu and Hillery, 1994; Prakash and Agarwal, 
1994, 1995; Gerry and Grobe, 1995; Puri and Agarwal, 1996; Luis and
Pe\v{r}ina, 1996; Brif and Ben-Aryeh, 1996; Brif and Mann 1996a,b), 
especially for generators of the SU(2) and SU(1,1) Lie groups. 
The SU(2) and SU(1,1) IS have been recently shown to be very useful 
for improving the accuracy of interferometric measurements 
(Hillery and Mlodinow, 1993; Brif and Ben-Aryeh, 1996; Brif and Mann 
1996a,b). 

The investigation of the AES yields the most full 
information on the IS for generators of the corresponding Lie group.
This information is of great importance in quantum optical 
applications of the IS. The most convenient way for examining 
different subsets of the AES and the relations between them is the 
construction of the analytic representation of the AES in the 
standard coherent-state basis. Actually, the idea to use such an 
analytic representation has been recently applied (Trifonov, 1994;
Brif and Ben-Aryeh, 1994a) to the SU(1,1) IS. As has been recently 
shown by Brif and Mann (1996a,b), the use of these representations 
is a powerful method for obtaining closed analytic expressions for 
various properties of the IS. In the present work the analytic 
representations are obtained for arbitrary AES of the SU(2) and 
SU(1,1) Lie groups. These representations are used for finding
the expansion of the AES in the corresponding orthonormal basis,
including the calculation of exact analytic expressions for the 
normalization factor and for some quantum expectation values.

\section{\uppercase{The SU(2) algebra eigenstates}}
\label{sec:SU2}
\setcounter{equation}{0}

In this section we discuss the AES for the SU(2) group which is the 
most elementary compact non-Abelian simple Lie group. The corresponding 
Lie algebra is spanned by the three operators 
$\{ J_{1} , J_{2} , J_{3} \}$,
\begin{equation}
[J_{1} , J_{2}] = iJ_{3} \ , \mbox{\hspace{0.8cm}} 
[J_{2} , J_{3}] = iJ_{1} \ , \mbox{\hspace{0.8cm}} 
[J_{3} , J_{1}] = iJ_{2} \ .   \label{4.1}
\end{equation}
It is convenient to use raising and lowering operators 
$J_{\pm}=J_{1} \pm iJ_{2}$ which satisfy the following commutation 
relations
\begin{equation}
[J_{3} , J_{\pm}] = \pm J_{\pm}, \mbox{\hspace{1.0cm}} 
[J_{-} , J_{+}] = -2 J_{3} \ .   \label{4.2}
\end{equation}
The Casimir operator
$J^{2} = J_{1}^{2} + J_{2}^{2} + J_{3}^{2}$
for any unitary irreducible representation is the identity operator 
times a number:
$J^{2} = j(j+1) I$.
Thus a representation of the SU(2) is determined by a single number 
$j$ that can be a positive integer or half-integer: 
$j=\frac{1}{2},1,\frac{3}{2},2,\ldots$. The representation Hilbert 
space is spanned by the orthonormal basis $|j,m\rangle$ 
($m=-j,-j+1,\ldots,j-1,j$).

\subsection{The standard coherent-state basis and related
analytic representation}

The standard set of the SU(2) CS is obtained for the lowest state 
$|j,-j\rangle$ chosen as the reference state. The isotropy subgroup 
$H$=U(1) consists of all group elements $h$ of the form 
$h=\exp(i\delta J_{3})$. Thus $h |j,-j\rangle = \exp(-i\delta j)
|j,-j\rangle$. The quotient space is SU(2)/U(1) (the sphere), and 
the standard coherent state is specified by a unit vector
\begin{equation}
\vec{n} = (\sin\theta\cos\varphi,\sin\theta\sin\varphi,\cos\theta).
\label{4.5}
\end{equation}
Then an element $\Omega$ of the quotient space can be written as 
\begin{equation}
\Omega \equiv D(\xi) = \exp (\xi J_{+} - \xi^{\ast} J_{-}) \ ,  
\label{4.6}
\end{equation}
where $\xi = -(\theta/2) e^{-i\varphi}$. The standard SU(2) CS are 
given by 
\begin{equation}
|j,\zeta\rangle = D(\xi) |j,-j\rangle = 
\exp(\xi J_{+} - \xi^{\ast} J_{-}) |j,-j\rangle = 
(1+|\zeta|^{2})^{-j} \exp(\zeta J_{+}) |j,-j\rangle \ ,   \label{4.7}
\end{equation}
where $\zeta = (\xi/|\xi|)\tan |\xi| = - \tan (\theta/2) 
e^{-i\varphi}$. The parameter $\zeta$ can acquire any complex 
value. The expansion of the $|j,\zeta\rangle$ states in the 
orthonormal basis is
\begin{equation}
|j,\zeta\rangle = (1+|\zeta|^{2})^{-j} \sum_{m=-j}^{j} \left[
\frac{(2j)!}{(j+m)!(j-m)!} \right]^{1/2} \zeta^{j+m} |j,m\rangle \ .
\label{4.8}  
\end{equation}  
The SU(2) CS are normalized but they are not orthogonal to each other:
\begin{equation}
\langle j,\zeta_{1}|j,\zeta_{2}\rangle = (1+|\zeta_{1}|^{2})^{-j} 
(1+|\zeta_{2}|^{2})^{-j} (1 + \zeta_{1}^{\ast}
\zeta_{2})^{2j} \ .   \label{4.9}
\end{equation}
The identity resolution is
\begin{equation}
\int d\mu(j,\zeta) |j,\zeta\rangle \langle j,\zeta| = I \ ,  
\mbox{\hspace{1.0cm}} d\mu(j,\zeta) = \frac{2j+1}{\pi}
\frac{d^{2}\!\zeta}{(1+|\zeta|^{2})^{2}} \ .  \label{4.10}
\end{equation}
For any state $| \Psi \rangle = \sum_{m=-j}^{j} c_{m} |j,m\rangle$ 
in the Hilbert space, one can construct the analytic function
\begin{equation}
f(\zeta) = (1+|\zeta|^{2})^{j} \langle j,\zeta^{\ast}|\Psi\rangle
 = \sum_{m=-j}^{j} c_{m} \left[\frac{(2j)!}{(j+m)!(j-m)!}
\right]^{1/2} \zeta^{j+m} \ .  \label{4.11}
\end{equation}
Then the state $| \Psi \rangle$ can be expanded in the standard 
coherent-state basis:
\begin{equation}
| \Psi \rangle = \int d\mu(j,\zeta) (1+|\zeta|^{2})^{-j} 
f(\zeta^{\ast}) |j,\zeta\rangle \ ,   \label{4.12}
\end{equation}
\begin{equation}
\langle\Psi|\Psi\rangle = \int d\mu(j,\zeta) (1+|\zeta|^{2})^{-2j} 
|f(\zeta^{\ast})|^{2}  < \infty \ .  \label{4.13}
\end{equation}
The coherent state $|j,\zeta_{0}\rangle$ is represented by the function
\begin{equation}
{\cal F}(j,\zeta_{0};\zeta) =  (1+|\zeta|^{2})^{j} 
\langle j,\zeta^{\ast}|j,\zeta_{0}\rangle = (1+|\zeta_{0}|^{2})^{-j}
(1+\zeta_{0}\zeta)^{2j} \ .   \label{4.14}
\end{equation}
The operators $J_{\pm}$ and $J_{3}$ act in the Hilbert space of 
analytic functions $f(\zeta)$ as first-order
differential operators
\begin{equation}
J_{+} = -\zeta^{2} \frac{d}{d\zeta} + 2j\zeta \ ,  
\mbox{\hspace{0.8cm}} J_{-} = \frac{d}{d\zeta} \ ,
\mbox{\hspace{0.8cm}} J_{3} = \zeta \frac{d}{d\zeta} -j \ .    
\label{4.15}
\end{equation}

\subsection{The general case}

The eigenvalue equation for the SU(2) AES is 
\begin{equation}
(\vec{\beta}\cdot \vec{J}) |j,\lambda,\vec{\beta}\rangle = 
(\beta_{1} J_{1} +\beta_{2} J_{2} + \beta_{3} J_{3}) 
|j,\lambda,\vec{\beta}\rangle
= \lambda |j,\lambda,\vec{\beta}\rangle \ .  \label{4.16}
\end{equation}
By introducing the analytic function
\begin{equation}
\Lambda(j,\lambda,\vec{\beta};\zeta) = (1+|\zeta|^{2})^{j} 
\langle j,\zeta^{\ast}|j,\lambda,\vec{\beta}\rangle \ ,  
\label{4.17}
\end{equation}
we derive the differential equation 
\begin{equation}
[\beta_{+} + \beta_{3} \zeta - \beta_{-} \zeta^{2} ] 
\frac{d}{d\zeta} \Lambda(j,\lambda,\vec{\beta};\zeta) + 
[ 2j \beta_{-} \zeta - j \beta_{3} -\lambda ] 
\Lambda(j,\lambda,\vec{\beta};\zeta) = 0 \ ,    \label{4.18}
\end{equation}
where we have defined
$\beta_{\pm} = ( \beta_{1} \pm i \beta_{2})/2$. Let us also define
\begin{equation}
b = \sqrt{ \beta_{1}^{2}+\beta_{2}^{2}+\beta_{3}^{2} } \ .  
\label{4.21}
\end{equation}
Admissible values of $\vec{\beta}$ and $\lambda$ are determined
by the requirement that the function 
$\Lambda(j,\lambda,\vec{\beta};\zeta)$ must be a polynomial of the
form (\ref{4.11}), i.e., it should be normalizable and analytic in 
the whole $\zeta$ plane. 
We will see that for any choice of $\vec{\beta}$, there exists at 
least one such a solution of equation (\ref{4.18}), i.e., each 
algebra element $(\vec{\beta}\cdot \vec{J})$ has at least one 
eigenstate in the Hilbert space of any irreducible representation 
of SU(2). In the general case $b \neq 0$, each algebra element has 
$2j+1$ eigenstates with symmetric spectrum: $\lambda = -jb, (-j+1)b,
\ldots, (j-1)b, jb$. In the degenerate case $b = 0$, each algebra 
element has only one eigenstate with eigenvalue $\lambda = 0$.
All these degenerate eigenstates are the standard CS. 

We start by considering the general case $b \neq 0$. 
For $\beta_{+} \neq 0$, the solution of equation (\ref{4.18}) reads
\begin{equation}
\Lambda(j,\lambda,\vec{\beta};\zeta) = {\cal N}^{-1/2} 
(1 - \tau_{-} \zeta)^{j+m_{0}}  (1 - \tau_{+} \zeta)^{j-m_{0}} \ ,  
\label{4.20}  \end{equation}
where ${\cal N}$ is a normalization factor, and we use the following
notation:
\begin{eqnarray}
& & \tau_{\pm} = (\beta_{1} - i\beta_{2})/(\beta_{3} \pm b) \ , \\
& & m_{0} = \lambda / b \ .
\end{eqnarray}
The condition of the analyticity for the function $\Lambda(\zeta)$ 
requires that $m_{0}$ can take only the values:
\begin{equation}
m_{0} = -j,-j+1,\ldots,j-1,j \ .    \label{4.22}
\end{equation}
This condition means that the SU(2) AES have the discrete spectrum
$\lambda = m_{0} b$.

We can compare the function $\Lambda(j,\lambda,\vec{\beta};\zeta)$ 
of equation (\ref{4.20}) with the function 
${\cal F}(j,\zeta_{0};\zeta)$ 
of equation (\ref{4.14}) that represents the standard coherent state 
$|j,\zeta_{0}\rangle$. We find that the algebra eigenstate 
$|j,\lambda,\vec{\beta}\rangle$ belongs to the standard set of the 
CS when $m_{0}=\pm j$. Then $\zeta_{0} = - \tau_{\mp}$,   
respectively. The normalization factor in this case is 
identified as ${\cal N} = (1+|\zeta_{0}|^{2})^{2j}$.
For example, we can choose $\vec{\beta}$ to be a unit vector 
$\vec{\beta} = \vec{n} = (\sin\theta \cos\varphi,\sin\theta 
\sin\varphi,\cos\theta)$, and $m_{0} = -j$. Then
\begin{equation}
\zeta_{0} = - \frac{ \sin\theta \, e^{-i\varphi} }{ \cos\theta +1 } = 
- \tan (\theta/2) \, e^{-i\varphi} \ .
\label{4.26}   
\end{equation}
It means that the standard CS form a subset of the AES with the 
corresponding eigenvalue equation
\begin{equation}
[ (\sin\theta \cos\varphi) J_{1} + (\sin\theta \sin\varphi) J_{2} + 
(\cos\theta) J_{3} ] |j,\zeta_{0}\rangle
= -j |j,\zeta_{0}\rangle \ .   \label{4.27}
\end{equation}
This result can be found by acting with $D(\xi_{0})$ on both 
sides of equation 
$J_{3} |j,-j\rangle = -j |j,-j\rangle$.

\subsection{The expansion in the orthonormal basis and
quantum statistics}
\label{su2expansion}

The decomposition of the AES $|j,\lambda,\vec{\beta}\rangle$ over 
the orthonormal basis can be obtained by expanding the function 
$\Lambda(j,\lambda,\vec{\beta};\zeta)$ of equation (\ref{4.20})
into a power series in $\zeta$. This can be done by using the
generating function for the Lagrange polynomials 
$g_{n}^{(\alpha,\beta)}(u,v)$ (Erd\'{e}lyi {\em et al.}, 1953, 
Vol.\ 3, Sec.\ 19.11; Srivastava and Manocha, 1984, Secs.\ 1.11, 8.5):
\begin{equation}
(1-u\zeta)^{-\alpha} (1-v\zeta)^{-\beta} = \sum_{n=0}^{\infty}
g_{n}^{(\alpha,\beta)}(u,v) \zeta^n \ .   \label{lpol-gf}
\end{equation}
The Lagrange polynomials are related to the more familiar Jacobi
polynomials $P_{n}^{(\alpha,\beta)}(x)$ via the relation 
(Srivastava and Manocha, 1984, Sec.\ 8.5)
\begin{equation}
g_{n}^{(\alpha,\beta)}(u,v) = (v-u)^{n} P_{n}^{(-\alpha-n,-\beta-n)}
\left( \frac{u+v}{u-v} \right) \ .  \label{polrel}
\end{equation}
Thus one obtains the following generating function for the
Jacobi polynomials:
\begin{equation}
(1-u\zeta)^{\mu} (1-v\zeta)^{\nu} = \sum_{n=0}^{\infty} (v-u)^{n} 
P_{n}^{(\mu-n,\nu-n)} \left( \frac{u+v}{u-v} \right) \zeta^n  \ .  
\label{jpol-gf}
\end{equation}
Using this expression, we obtain the power series for the function 
$\Lambda(j,\lambda,\vec{\beta};\zeta)$ of equation (\ref{4.20}):
\begin{equation}
\Lambda(j,\lambda,\vec{\beta};\zeta) 
= {\cal N}^{-1/2} \sum_{m=-j}^{j} P_{j+m}^{(m_{0}-m,-m_{0}-m)}(x)\, 
(\kappa \zeta)^{j+m} \ , \label{pser2}
\end{equation}
where we have defined
\begin{eqnarray}
& & \kappa = \tau_{+} - \tau_{-} = 2b/(\beta_{1}+i\beta_{2}) \ , \\
& & x = (\tau_{-} + \tau_{+})/(\tau_{-} - \tau_{+}) = \beta_{3}/b \ .
\end{eqnarray}
The series (\ref{pser2}) is finite due to the fact that
\begin{equation}
P_{n}^{(j+m_{0}-n,j-m_{0}-n)}(x) = 0 \;\;\;\; {\rm for} \;
n > 2j \ .  \label{n2j}
\end{equation}
Comparing the expansion (\ref{pser2}) with the general formula 
(\ref{4.11}), we find the decomposition of the AES over the 
orthonormal basis:
\begin{equation}
|j,\lambda,\vec{\beta}\rangle = {\cal N}^{-1/2} \sum_{m=-j}^{j}
\left[ \frac{ (j+m)! (j-m)! }{ (2j)! } \right]^{1/2}
P_{j+m}^{(m_{0}-m,-m_{0}-m)}(x)\, \kappa^{j+m} |j,m\rangle \ .
\label{decomp2}
\end{equation}

The normalization factor is given by
\begin{equation}
{\cal N} = \sum_{n=0}^{2j} \frac{ n! (2j-n)! }{ (2j)! }
\left| P_{n}^{(j+m_{0}-n,j-m_{0}-n)}(x) \right|^2 t^{n} \ , 
\label{Nseries}
\end{equation}
where $t = |\kappa|^2$. The summation in (\ref{Nseries}) can be
formally continued up to infinity because all the terms with
$n > 2j$ vanish. Then we can use the summation theorem for the
Jacobi polynomials (Srivastava and Manocha, 1984, Sec.\ 2.3, Eqs.\
60, 62), that can be written in the form
\begin{equation}
\sum_{n=0}^{\infty} \frac{ n! \Gamma(\mu+\nu+1-n) 
}{ \Gamma(\mu+\nu+1) } \left| P_{n}^{(\mu-n,\nu-n)}(x)
\right|^2 t^{n} = S_{+}^{\mu} S_{-}^{\nu} 
F\left(-\nu,-\mu;-\mu-\nu;\frac{t}{S_{+}S_{-}} \right) \ ,   
\label{jp-sum}
\end{equation}
for $\mu+\nu \geq 0$. Here, $F(a,b;c;z)$ is the hypergeometric 
function, and we have defined
\begin{equation}
S_{\pm} = 1 + |x \pm 1|^{2} t/4 = 1 + |\tau_{\mp}|^{2} \ .
\end{equation}
If $\mu$ and $\nu$ are nonnegative integers, we can use the 
relation between the hypergeometric function and the Jacobi 
polynomials (Erd\'{e}lyi {\em et al.}, 1953, Vol.\ 2, Sec.\ 10.8), 
that can be expressed in the form
\begin{equation}
F(-\mu,-\nu;-\mu-\nu;z) = (-1)^{n} \frac{ \mu! \nu! }{ (\mu+\nu)! } 
P_{n}^{(-\mu-\nu-1,0)}(1-2z) \ ,  \;\;\;\;\;\; n = \min(\mu,\nu) \ .
\label{hf-jpol-rel}
\end{equation}
Therefore, we obtain the closed expression for the normalization 
factor:
\begin{equation}
{\cal N} = (-1)^{j-|m_{0}|} S_{+}^{j+m_{0}} S_{-}^{j-m_{0}}
\frac{ (j+m_{0})! (j-m_{0})! }{ (2j)! } P_{j-|m_{0}|}^{(-2j-1,0)}
\left( 1-\frac{2t}{S_{+}S_{-}} \right) \ .
\end{equation}
It can be easily verified that for $m_{0} = \pm j$, these formulas
reduce to the corresponding results for the standard coherent state 
$|j,\zeta_{0}\rangle$ with $\zeta_{0} = -\tau_{\mp}$, respectively.

The above analytic expressions can be used for calculations of
quantum statistical properties of the SU(2) AES. We demonstrate
how such a calculation can be performed by considering moments
of the generator $J_{3}$. By using the property 
$J_{3} |j,m\rangle = m |j,m\rangle$ and formula (\ref{Nseries}),
we can express moments of $J_{3}$ over the AES as derivatives of
${\cal N}$ with respect to $t$:
\begin{equation}
\langle J_{3} \rangle = \frac{t}{{\cal N}} 
\frac{\partial {\cal N}}{\partial t} - j \ , 
\end{equation}
\begin{equation}
(\Delta J_{3})^{2} = \frac{t^{2}}{{\cal N}} 
\frac{\partial^{2} {\cal N}}{\partial t^{2}} +
\frac{t}{{\cal N}} \frac{\partial {\cal N}}{\partial t}
- \left( \frac{t}{{\cal N}} \frac{\partial {\cal N}}{\partial t}
\right)^{2} \ . 
\end{equation}
By using the formula (Erd\'{e}lyi {\em et al.}, 1953, Vol.\ 2, 
Sec.\ 10.8)
\begin{equation}
\frac{ d P_{n}^{(\mu,\nu)}(x) }{ d x } = \frac{ n+\mu+\nu+1 }{2}
P_{n-1}^{(\mu+1,\nu+1)}(x)
\end{equation}
and the differential equation for the Jacobi polynomials, we obtain
exact analytic expressions for the moments of $J_{3}$:
\begin{equation}
\langle J_{3} \rangle = \frac{j Y + m_{0}(S_{+}-S_{-})}{ S_{+}S_{-} }
- \frac{ (j+|m_{0}|) Y t }{ S_{+}^{2} S_{-}^{2} } \Omega \ ,
\end{equation}
\begin{eqnarray}
(\Delta J_{3})^{2} & = & (j+m_{0}) \frac{ S_{+}-1 }{ S_{+}^{2} }
+ (j-m_{0}) \frac{ S_{-}-1 }{ S_{-}^{2} } + \frac{ (j^{2}-m_{0}^{2})
Y^{2} t }{ ( S_{+}S_{-} - t) S_{+}^{2} S_{-}^{2} } \nonumber \\
& & + \frac{ (j+|m_{0}|)\, t }{ S_{+}^{3} S_{-}^{3} }
\left( \frac{ S_{+} S_{-} Y^{2} }{ S_{+}S_{-} - t } + 2 j Y^2 
+ Z \right) \Omega  
- \frac{ (j+|m_{0}|)^{2} Y^{2} t^{2} }{ S_{+}^{4} S_{-}^{4} }\,
\Omega^{2} \ .
\end{eqnarray}
Here, we have introduced the following notation:
\begin{equation}
Y = S_{+} S_{-} - S_{+} - S_{-}  \ , \label{Ydef}
\end{equation}
\begin{equation}
Z = S_{+}^{2}(1- S_{-}) + S_{-}^{2}(1 - S_{+})  \ ,  \label{Zdef}
\end{equation}
\begin{equation}
\Omega = \left[ P_{j-|m_{0}|}^{(-2j-1,0)} 
\left(1-\frac{2t}{S_{+}S_{-}} \right) \right]^{-1}
P_{j-|m_{0}|-1}^{(-2j,1)} \left(1-\frac{2t}{S_{+}S_{-}} \right) \ ,
\;\;\;\;\; |m_{0}| < j \ .
\end{equation}
For $m_{0} = \pm j$, we have $\Omega = 0$, and then we recover the 
known results for the SU(2) CS (Wodkiewicz and Eberly, 1985).
The expressions for $\langle J_{3} \rangle$ and $(\Delta J_{3})^{2}$ 
are significantly simplified in the case $Y = 0$, which means 
\begin{equation}
|\tau_{+}\tau_{-}| = 1 \; \Leftrightarrow \; \left| \frac{
(\beta_{1} - i \beta_{2})^2 }{ \beta_{1}^{2} + \beta_{2}^{2} }
\right| = 1 \ .    \label{Y0cond}
\end{equation}
This condition is satisfied in the important case 
$\beta_{1} = a \beta_{2}$, where $a$ is any real number
(this includes the case when $\beta_{1}$ or $\beta_{2}$
vanishes). Then we obtain
\begin{equation}
\langle J_{3} \rangle = \frac{h-1}{h+1}  m_{0} \ , \label{j3ms}
\end{equation}
\begin{equation}
(\Delta J_{3})^{2} = \frac{ 2 j h }{ (h+1)^{2} } - 
\frac{ 2 (j+|m_{0}|) h^2 t }{ (h+1)^{4} }\, \Omega \ , \label{j3vs}
\end{equation}
where
\begin{equation}
h = |\tau_{-}|^2 = 1/|\tau_{+}|^2 \ .  \label{hdef}
\end{equation}

\subsection{Some special cases}

For $\beta_{+} = 0$ and $\beta_{3} \neq 0$, we obtain $\tau_{-} 
\rightarrow \infty$, so we cannot use formula (\ref{4.20}). 
In this case the solution of equation (\ref{4.18}) is 
\begin{equation}
\Lambda(j,\lambda,\vec{\beta};\zeta) = {\cal N}^{-1/2} 
\zeta^{j+m_{0}} (1 - \tau_{+} \zeta )^{j-m_{0}}  \label{lam-spec} \ ,
\end{equation}
where $\tau_{+} = \beta_{1}/\beta_{3}$ and $m_{0}=\lambda/\beta_{3}$. 
The condition of the analyticity requires that $m_{0}$ can take only 
the discrete values (\ref{4.22}). The results of section 
\ref{su2expansion} cannot be used in this special case, but 
corresponding expressions can be obtained by expanding the function 
(\ref{lam-spec}) into the power series. We find
\begin{equation}
|j,\lambda,\vec{\beta}\rangle = {\cal N}^{-1/2} \sum_{m=m_{0}}^{j} 
\left[ \frac{ (j+m)! }{ (j-m)! } \right]^{1/2} \frac{ 
(-\tau_{+})^{m-m_{0}} }{ (m-m_{0})! } |j,m\rangle \ ,
\end{equation}
\begin{equation}
{\cal N} = \frac{ (j+m_{0})! }{ (j-m_{0})! }\, 
P_{j-m_{0}}^{(0,2m_{0})} \left( 2|\tau_{+}|^{2} + 1 \right) \ .
\end{equation}
For $m_{0}=-j$, the function (\ref{lam-spec}) represents the standard 
coherent state $|j,\zeta_{0}\rangle$ with $\zeta_{0} = - \tau_{+}$. 
The corresponding eigenvalue equation is 
\begin{equation}
( J_{3} - \zeta_{0} J_{+} ) |j,\zeta_{0}\rangle = 
-j |j,\zeta_{0}\rangle \ .
\end{equation}
For $m_{0}=j$, we find $\Lambda(\zeta) = \zeta^{2j}$ that represents 
the state $|j,j\rangle$.

For $\beta_{-} = 0$ and $\beta_{3} \neq 0$, we can use the general
results of the preceding sections with $b = \beta_{3}$, $\tau_{+} = 0$ 
and $\tau_{-} = -\beta_{3}/\beta_{1}$. This gives $\kappa = -\tau_{-}
= \beta_{3}/\beta_{1}$, $x = 1$, $S_{+} = 1 + t$ (where 
$t = |\kappa|^{2}$) and $S_{-} = 1$. The corresponding analytic 
function is
\begin{equation}
\Lambda(j,\lambda,\vec{\beta};\zeta) = {\cal N}^{-1/2}
(1 - \tau_{-} \zeta )^{j+m_{0}} \ .
\end{equation}
For $m_{0} = j$, this function represents the standard coherent
state $|j,\zeta_{0}\rangle$ with $\zeta_{0} = \beta_{3}/\beta_{1}$.
The corresponding eigenvalue equation is 
\begin{equation}
( J_{3} + \zeta_{0}^{-1} J_{-} ) |j,\zeta_{0}\rangle = 
j |j,\zeta_{0}\rangle \ .
\end{equation}
For $m_{0} = -j$, we find $\Lambda(\zeta) = 1$ that corresponds
to the state $|j,-j\rangle$.

For the degenerate case $b=0$, the solution of equation (\ref{4.18}) 
is 
\begin{equation}
\Lambda(j,\lambda,\vec{\beta};\zeta) = {\cal N}^{-1/2} 
( 1 - \tau \zeta )^{2j} \exp\left( - \frac{ 2\lambda }{ \beta_{3} }
\frac{ 1 }{ 1 - \tau \zeta } \right)   \ ,   
\end{equation}
where $\tau = 2\beta_{-}/\beta_{3} = -\beta_{3}/2\beta_{+}$.
This function is analytic only for $\lambda =0$. Then we obtain
\begin{equation}
\Lambda(j,\lambda=0,\vec{\beta};\zeta) = {\cal N}^{-1/2} 
( 1 - \tau \zeta )^{2j} \ .
\label{4.32}   
\end{equation}
This function represents the standard coherent state 
$|j,\zeta_{0}\rangle$ with $\zeta_{0} = -\tau$ and
${\cal N} = (1+|\zeta_{0}|^{2})^{2j}$.
For example, we can choose $\vec{\beta} = 
\Bigl(1-\zeta_{0}^{2}, -i(1+\zeta_{0}^{2}), 2\zeta_{0} \Bigr)$. 
Then the standard CS satisfy the eigenvalue equation
\begin{equation}
( J_{-} + 2 \zeta_{0} J_{3} - \zeta_{0}^{2} J_{+} ) 
|j,\zeta_{0}\rangle = 0 \ .   \label{4.34}
\end{equation}
For $\beta_{-}=\beta_{3}=0$, the only normalizable solution 
of equation (\ref{4.18}) is $\Lambda(\zeta) = 1$  
that represents the state $|j,-j\rangle$ (that is the standard 
coherent state with $\zeta_{0} = 0$). 
Analogously, for $\beta_{+}=\beta_{3}=0$, the corresponding function 
is $\Lambda(\zeta) = \zeta^{2j}$ that represents 
the state $|j,j\rangle$ (that is the standard coherent state with 
$\zeta_{0} \rightarrow \infty$).

\subsection{The SU(2) intelligent states}

The SU(2) generalized IS were defined by Trifonov (1994) 
as the eigenstates of the operator $\eta J_{1} -iJ_{2}$
[see equation (\ref{2.18})]. In our notation, the generalized IS 
are the AES with $\vec{\beta} = (\eta,-i,0)$. For $\eta^{2} \neq 1$, 
the corresponding analytic function is given by the particular case 
of equation (\ref{4.20}) with $b=\sqrt{\eta^2-1}$ and
\begin{equation}
\tau_{\pm} = \pm \sqrt{ \frac{ \eta -1 }{ \eta +1 } } \ .
\end{equation}
All the results of section \ref{su2expansion} are valid here with 
$\kappa = 2\tau_{+}$, $x=0$, and $S_{+}=S_{-} = 1+|\tau_{+}|^{2}$.
The generalized intelligent state is also the standard coherent 
state when $m_{0} = \pm j$. The corresponding coherent-state amplitude 
is $\zeta_{0} =  -\tau_{\mp}$, respectively. Since $\eta$ is an 
arbitrary complex number (excluding $\pm 1$), $\zeta_{0}$ also can 
acquire any complex value (excluding $0$ and $\infty$). For $\eta =1$, 
the generalized IS are reduced to the state $|j,-j\rangle$ 
that is the standard coherent state with $\zeta_{0} = 0$. For 
$\eta =-1$, the generalized IS are reduced to the state $|j,j\rangle$ 
that is the standard coherent state with 
$\zeta_{0} \rightarrow \infty$. Hence, the standard set of the 
SU(2) CS is a subset of the SU(2) generalized IS.

We also discuss briefly three types of the SU(2) ordinary IS. The 
$J_{1}$-$J_{2}$ IS are determined, according to equation 
(\ref{2.19}), as the eigenstates 
of the operator $J_{1} +i\gamma J_{2}$, where $\gamma$ is a real 
parameter. These states are the SU(2) AES with $\vec{\beta} = 
(1,i\gamma,0)$. For $\gamma^{2} \neq 1$, the corresponding analytic 
function is given by the particular case of equation (\ref{4.20})
with $b=\sqrt{1-\gamma^{2}}$ and
\begin{equation}
\tau_{\pm} = \pm \sqrt{ \frac{ 1+\gamma }{ 1-\gamma } } \ .
\end{equation}
All the results of section \ref{su2expansion} are valid here with 
$\kappa = 2\tau_{+}$, $x=0$, and $S_{+}=S_{-} = 1+|\tau_{+}|^{2}$.
The $J_{1}$-$J_{2}$ IS coincide with the standard CS for 
$m_{0} = \pm j$. Since $\gamma$ is real, the corresponding 
coherent-state amplitude $\zeta_{0}$ is real for 
$|\gamma| < 1$ and pure imaginary for $|\gamma| > 1$.
Therefore the set of the $J_{1}$-$J_{2}$ IS and the standard set of 
the SU(2) CS have an intersection.

The $J_{1}$-$J_{3}$ IS are determined, according to equation 
(\ref{2.19}), as the eigenstates of the operator 
$J_{1} +i\gamma J_{3}$. These states are the SU(2) AES with 
$\vec{\beta} = (1,0,i\gamma)$. For $\gamma^{2} \neq 1$, the 
corresponding analytic function is given by the particular case of 
equation (\ref{4.20}) with $b=\sqrt{1-\gamma^{2}}$ and
\begin{equation}
\tau_{\pm} = \left( i \gamma \pm \sqrt{1-\gamma^{2}} \right)^{-1} \ .
\end{equation}
All the results of section \ref{su2expansion} are valid here with 
$\kappa = 2\sqrt{1-\gamma^{2}}$, $x= i \gamma/\sqrt{1-\gamma^{2}}$.
Note that $|\tau_{+}\tau_{-}| = 1$, and therefore we can use simple
expressions (\ref{j3ms}) and (\ref{j3vs}). For $\gamma^{2} < 1$,
we have $S_{+}=S_{-}= 2$ and $h=1$. For $\gamma^{2} > 1$, we
obtain $h = 2\gamma^{2}+2\gamma\sqrt{\gamma^{2}-1}-1$.
The intersection between the $J_{1}$-$J_{3}$ IS and the standard 
CS is obtained for $m_{0} = \pm j$. In the case $\gamma = \pm 1$, 
we have $\lambda = b = 0$, and then the corresponding analytic 
function is $\Lambda(\zeta) = {\cal N}^{-1/2} (1 \pm i \zeta)^{2j}$.  
This function corresponds to the standard coherent state 
$|j,\zeta_{0}\rangle$ with $\zeta_{0} = \pm i$.

The $J_{2}$-$J_{3}$ IS are determined, according to equation 
(\ref{2.19}), as the eigenstates of the operator 
$J_{2} +i\gamma J_{3}$. These states are the SU(2) AES with 
$\vec{\beta} = (0,1,i\gamma)$. For $\gamma^{2} \neq 1$, the 
corresponding analytic function is given by the particular case of 
equation (\ref{4.20}) with $b=\sqrt{1-\gamma^{2}}$ and
\begin{equation}
\tau_{\pm} = \left( - \gamma \pm i \sqrt{1-\gamma^{2}} \right)^{-1} \ .
\end{equation}
All the results of section \ref{su2expansion} are valid here with 
$\kappa = -2 i \sqrt{1-\gamma^{2}}$, $x= i \gamma/\sqrt{1-\gamma^{2}}$.
Here $|\tau_{+}\tau_{-}| = 1$, and therefore we can use simple
expressions (\ref{j3ms}) and (\ref{j3vs}). For $\gamma^{2} < 1$,
we find $S_{+}=S_{-}= 2$ and $h=1$. For $\gamma^{2} > 1$, we
obtain $h = 2\gamma^{2}+2\gamma\sqrt{\gamma^{2}-1}-1$.
The intersection between the $J_{1}$-$J_{3}$ IS and the standard 
CS is obtained for $m_{0} = \pm j$. In the case $\gamma = \pm 1$, 
we have $\lambda = b = 0$, and then the corresponding analytic 
function is $\Lambda(\zeta) = {\cal N}^{-1/2} (1 \pm \zeta)^{2j}$.  
This function corresponds to the standard coherent state 
$|j,\zeta_{0}\rangle$ with $\zeta_{0} = \pm 1$.

\section{\uppercase{The SU(1,1) algebra eigenstates}}
\label{sec:SU1}
\setcounter{equation}{0}

In this section we consider the AES for the SU(1,1) group which is 
the most elementary noncompact non-Abelian simple Lie group. It has 
several series of unitary irreducible representations: discrete, 
continuous and supplementary (Bargmann, 1947; Vilenkin, 1968). In 
the present work we discuss only the case of the discrete series 
which has many well-known physical applications (Perelomov, 1986). 
The Lie algebra corresponding to the group SU(1,1) is spanned by the 
three operators $\{ K_{1} , K_{2} , K_{3} \}$,
\begin{equation}
[K_{1} , K_{2}] = -iK_{3} \ , \mbox{\hspace{0.8cm}} 
[K_{2} , K_{3}] = iK_{1} \ , \mbox{\hspace{0.8cm}} 
[K_{3} , K_{1}] = iK_{2} \ .   \label{5.1}
\end{equation} 
It is convenient to use raising and lowering operators 
$K_{\pm}=K_{1} \pm iK_{2}$ which satisfy the following
commutation relations:
\begin{equation}
[K_{3} , K_{\pm}] = \pm K_{\pm}, \mbox{\hspace{1.0cm}} 
[K_{-} , K_{+}] = 2 K_{3} \ .   \label{5.2}
\end{equation}
The Casimir operator
$K^{2} = K_{3}^{2} - K_{1}^{2} - K_{2}^{2}$
for any unitary irreducible representation is the identity operator 
times a number:
$K^{2} = k(k-1) I$. Representations of SU(1,1) are
determined by a single number $k$; for the discrete-series 
representations this number acquires discrete values 
$k=\frac{1}{2},1,\frac{3}{2},2,\ldots$. 
The representation Hilbert space is spanned by the orthonormal basis 
$|k,n\rangle$ ($n=0,1,2,\ldots$).

The SU(1,1) AES can be investigated by using two alternative 
analytic representations: one of them is based on the standard 
CS (Perelomov, 1986) and the other is based on the so-called 
Barut-Girardello (BG) states (Barut and Girardello, 1971).

\subsection{The standard coherent-state basis and the 
analytic representation in the unit disk}

The standard set of the SU(1,1) CS is obtained for the lowest state 
$|k,0\rangle$ chosen as the reference state. The isotropy subgroup 
$H$=U(1) consists of all group elements $h$ of the form 
$h=\exp(i\delta K_{3})$. Thus $h|k,0\rangle = 
\exp(i\delta k)|k,0\rangle$. The quotient space is SU(1,1)/U(1) 
(the upper sheet of the two-sheet hyperboloid), and the standard 
coherent state is specified by a unit pseudo-Euclidean vector
\begin{equation}
\vec{n} = (\sinh\chi \cos\varphi,\sinh\chi \sin\varphi,\cosh\chi) \ .  
\label{5.5}
\end{equation}
Then an element $\Omega$ of the quotient space can be written as 
\begin{equation}
\Omega \equiv D(\xi) = \exp (\xi K_{+} - \xi^{\ast} K_{-}) \ ,  
\label{5.6}
\end{equation}
where $\xi = -(\chi/2) e^{-i\varphi}$. The standard SU(1,1) CS are 
given by 
\begin{equation}
|k,\zeta\rangle = D(\xi) |k,0\rangle = \exp(\xi K_{+} - \xi^{\ast} 
K_{-}) |k,0\rangle = (1-|\zeta|^{2})^{k}
\exp(\zeta K_{+}) |k,0\rangle \ ,   \label{5.7}
\end{equation}
where $\zeta = (\xi/|\xi|)\tanh |\xi| = - \tanh (\chi/2) 
e^{-i\varphi}$. The parameter $\zeta$ is restricted by $|\zeta|<1$. 
The expansion of the $|k,\zeta\rangle$ states in the orthonormal 
basis is
\begin{equation}
|k,\zeta\rangle = (1-|\zeta|^{2})^{k} \sum_{n=0}^{\infty} 
\left[\frac{\Gamma(2k+n)}{n!\Gamma(2k)}\right]^{1/2} 
\zeta^{n} |k,n\rangle \ .
\label{5.8}  \end{equation}  
The SU(1,1) CS are normalized but they are not orthogonal to each 
other:
\begin{equation}
\langle k,\zeta_{1}|k,\zeta_{2}\rangle = (1-|\zeta_{1}|^{2})^{k} 
(1-|\zeta_{2}|^{2})^{k} (1 - \zeta_{1}^{\ast}
\zeta_{2})^{-2k} \ .   \label{5.9}
\end{equation}
The identity resolution is (for $k > \frac{1}{2}$)
\begin{equation}
\int d\mu(k,\zeta) |k,\zeta\rangle \langle k,\zeta| = I \ ,  
\mbox{\hspace{1.0cm}} d\mu(k,\zeta) = \frac{2k-1}{\pi}
\frac{d^{2}\!\zeta}{(1-|\zeta|^{2})^{2}} \ ,  \label{5.10}
\end{equation}
and for $k=\frac{1}{2}$ the limit $k\rightarrow \frac{1}{2}$ must be 
taken after the integration is carried out in the general form. For 
any state $| \Psi \rangle = \sum_{n=0}^{\infty} c_{n} |k,n\rangle$ 
in the Hilbert space, one can construct the analytic function
\begin{equation}
f(\zeta) = (1-|\zeta|^{2})^{-k} \langle k,\zeta^{\ast}|\Psi\rangle 
= \sum_{n=0}^{\infty} c_{n} 
\left[\frac{\Gamma(2k+n)}{n!\Gamma(2k)}\right]^{1/2} \zeta^{n} \ .  
\label{5.11}
\end{equation}
Since $|\zeta|<1$, this analytic representation is referred to as 
the representation in the unit disk. The expansion of the state 
$| \Psi \rangle$ in the standard coherent-state basis is given by
\begin{equation}
| \Psi \rangle = \int d\mu(k,\zeta) (1-|\zeta|^{2})^{k} 
f(\zeta^{\ast}) |k,\zeta\rangle \ ,   \label{5.12}
\end{equation}
\begin{equation}
\langle\Psi|\Psi\rangle = \int d\mu(k,\zeta) (1-|\zeta|^{2})^{2k} 
|f(\zeta^{\ast})|^{2}  < \infty \ .  \label{5.13}
\end{equation}
The coherent state $|k,\zeta_{0}\rangle$ is represented by the 
function
\begin{equation}
{\cal F}(k,\zeta_{0};\zeta) =  (1-|\zeta|^{2})^{-k} \langle 
k,\zeta^{\ast}|k,\zeta_{0}\rangle = (1-|\zeta_{0}|^{2})^{k}
(1-\zeta_{0}\zeta)^{-2k} \ .   \label{5.14}
\end{equation}
The operators $K_{\pm}$ and $K_{3}$ act in the Hilbert space of 
analytic functions $f(\zeta)$ as first-order differential operators 
\begin{equation}
K_{+} = \zeta^{2} \frac{d}{d\zeta} + 2k\zeta \ ,  
\mbox{\hspace{0.8cm}} K_{-} = \frac{d}{d\zeta} \ ,
\mbox{\hspace{0.8cm}} K_{3} = \zeta \frac{d}{d\zeta} +k \ .    
\label{5.15}
\end{equation}

\subsection{The general case}
\label{su11general}

The eigenvalue equation for the SU(1,1) AES is 
\begin{equation}
(\vec{\beta}\cdot \vec{K}) |k,\lambda,\vec{\beta}\rangle = 
(\beta_{1} K_{1} +\beta_{2} K_{2} + \beta_{3} K_{3}) 
|k,\lambda,\vec{\beta}\rangle
= \lambda |k,\lambda,\vec{\beta}\rangle \ .  \label{5.16}
\end{equation}
Some particular cases of this equation were considered by Barut and
Girardello (1971), Lindblad and Nagel (1970), Solomon (1971) and
Nagel (1995). 
By introducing the analytic function
\begin{equation}
\Lambda(k,\lambda,\vec{\beta};\zeta) = (1-|\zeta|^{2})^{-k} 
\langle k,\zeta^{\ast}|k,\lambda,\vec{\beta}\rangle \ ,  \label{5.17}
\end{equation}
we derive the differential equation 
\begin{equation}
[\beta_{+} + \beta_{3} \zeta + \beta_{-} \zeta^{2} ] 
\frac{d}{d\zeta} \Lambda(k,\lambda,\vec{\beta};\zeta) 
+ [ 2k \beta_{-} \zeta + k \beta_{3} -\lambda ] 
\Lambda(k,\lambda,\vec{\beta};\zeta) = 0 \ ,    \label{5.18}
\end{equation}
where we have defined
$\beta_{\pm} = ( \beta_{1} \pm i \beta_{2})/2$. 
Let us also define
\begin{equation}
b = \sqrt{ \beta_{3}^{2}-\beta_{1}^{2}-\beta_{2}^{2} }  \ .  
\label{5.21}
\end{equation}
Admissible values of $\vec{\beta}$ and $\lambda$ are determined by
the requirement that the function 
$\Lambda(k,\lambda,\vec{\beta};\zeta)$ must be analytic in the unit 
disk. We will see that the non-compactness of the SU(1,1) group leads
to a reach structure that is absent in the SU(2) case. In the
general case $b \neq 0$, there are three classes of algebra elements
$(\vec{\beta} \cdot \vec{K})$. The first class consists of elements 
with a continuous spectrum (no restrictions on $\lambda$). Elements 
in the second class have a discrete equidistant spectrum: in one 
subclass $\lambda = (k+l)b$ and in the other subclass $\lambda = 
-(k+l)b$ (where $l=0,1,2,\ldots$). The third class includes elements 
that have not any normalizable eigenstate. In the degenerate case 
$b = 0$, there are two classes of algebra elements. The first class 
consists of elements with a continuous spectrum (no restrictions on 
$\lambda$), while the second class includes elements that have not 
any normalizable eigenstate. In the degenerate case there are no 
algebra elements with a discrete spectrum.

We first consider the general case $b \neq 0$. For $\beta_{+} \neq 0$,
the solution of equation (\ref{5.18}) is 
\begin{equation}
\Lambda(k,\lambda,\vec{\beta};\zeta) = {\cal N}^{-1/2} 
( 1 + \tau_{-} \zeta )^{-k+r} ( 1 + \tau_{+} \zeta)^{-k-r} \ ,  
\label{5.20}  
\end{equation}
where ${\cal N}$ is a normalization factor, and we use the
following notation:
\begin{eqnarray}
\label{taupm-def}
& & \tau_{\pm} = (\beta_{1} - i \beta_{2})/(\beta_{3} \pm b) \, \\
& & r = \lambda/b  \ .  \label{r-def} 
\end{eqnarray}
Now we analyse the analyticity condition for the function 
$\Lambda(\zeta)$ of equation (\ref{5.20}). 
If $|\tau_{+}| < 1$ and $|\tau_{-}| < 1$,
then there are no restrictions on $\lambda$ (i.e., the corresponding
algebra elements have a continuous complex spectrum). 
If $|\tau_{+}| < 1$ and $|\tau_{-}| \geq 1$, then the analyticity 
condition requires $r = k+l$ (where $l=0,1,2,\ldots$), i.e., 
the spectrum is discrete and equidistant:
\begin{equation}
\lambda = (k+l)b \ .
\end{equation}
If $|\tau_{+}| \geq 1$ and $|\tau_{-}| < 1$, then the analyticity 
condition requires $r = -(k+l)$, and once again the spectrum is 
discrete and equidistant:
\begin{equation}
\lambda = -(k+l)b \ .
\end{equation}
If $|\tau_{+}| \geq 1$ and $|\tau_{-}| \geq 1$, then the function 
$\Lambda(\zeta)$ of equation (\ref{5.20}) cannot be analytic in
the unit disk for any value of $\lambda$. This region in the 
parameter space is forbidden, i.e., the corresponding algebra
elements have no normalizable eigenstates. Note that there are 
algebra elements whose eigenstates are unnormalizable in sense of
equation (\ref{5.13}), but these states can be orthonormalized by
the delta function. For example, such generalized orthonormality
relations exist for the eigenstates of the operators $K_{1}$ and
$K_{2}$, for which $|\tau_{+}| = |\tau_{-}| = 1$ (Lindblad and
Nagel, 1970; Nagel, 1995). In our notation, these operators 
formally belong to the forbidden region of the parameter space.
The structure of the parameter space is described in Fig.~1. 
%%%%%%%%%%%%%%%%%%%
\begin{figure}[htbp]
%\vspace*{-1cm}
%\epsfxsize=0.7\textwidth
%\centerline{\epsffile{su11aes.ps}}\vspace*{0.2cm}
\caption{The structure of the parameter space for the SU(1,1) AES
in the general case $b \neq 0$.}
\end{figure}
%%%%%%%%%%%%%%%%%%%

We can compare the function $\Lambda(k,\lambda,\vec{\beta};\zeta)$ 
of equation (\ref{5.20}) with the function 
${\cal F}(k,\zeta_{0};\zeta)$ 
of equation (\ref{5.14}) that represents the standard coherent state 
$|k,\zeta_{0}\rangle$. We find that the algebra eigenstate 
$|k,\lambda,\vec{\beta}\rangle$ belongs to the standard set of the 
CS when $r = \pm k$, i.e., $\lambda = \pm k b$. Then
$\zeta_{0} = - \tau_{\pm}$, respectively. The condition 
$|\zeta_{0}|<1$ is equivalent to the analyticity condition:
$|\tau_{\pm}|<1$ for $r = \pm k$, respectively.
The normalization factor ${\cal N}$ in this case is identified as
${\cal N} = ( 1-|\zeta_{0}|^{2} )^{-2k}$. For example, we can 
choose $\vec{\beta}$ to be a unit pseudo-Euclidean vector 
$\vec{\beta} = \vec{n} = (\sinh\chi \cos\varphi,\sinh\chi 
\sin\varphi,\cosh\chi)$, and $r = k$. Then
\begin{equation}
\zeta_{0} = - \frac{ \sinh\chi \, e^{-i\varphi} }{ \cosh\chi +1 } = 
- \tanh (\chi/2) \, e^{-i\varphi} \ .
\label{5.26}   
\end{equation}
It means that the standard CS form a subset of the AES with the 
corresponding eigenvalue equation
\begin{equation}
[ (\sinh\chi \cos\varphi) K_{1} + (\sinh\chi \sin\varphi) K_{2} + 
(\cosh\chi) K_{3} ] |k,\zeta_{0}\rangle
= k |k,\zeta_{0}\rangle \ .   \label{5.27}
\end{equation}
This result can be found by acting with $D(\xi_{0})$ on both 
sides of equation 
$K_{3}|k,0\rangle=k|k,0\rangle$.

\subsection{The expansion in the orthonormal basis 
and quantum statistics}
\label{su11expansion}

In the allowed region of the parameter space, we can use equation 
(\ref{jpol-gf}) for expanding the function 
$\Lambda(k,\lambda,\vec{\beta};\zeta)$ of equation (\ref{5.20})
into the power series:
\begin{equation}
\Lambda(k,\lambda,\vec{\beta};\zeta) 
= {\cal N}^{-1/2} \sum_{n=0}^{\infty} P_{n}^{(-k+r-n,-k-r-n)}(x)\, 
(\kappa \zeta)^{n} \ , \label{pser11}
\end{equation}
where we have defined
\begin{eqnarray}
& & \kappa = \tau_{+} - \tau_{-} = -2b/(\beta_{1}+i\beta_{2}) \ , \\
& & x = (\tau_{-} + \tau_{+})/(\tau_{-} - \tau_{+}) = \beta_{3}/b \ .
\end{eqnarray}
Comparing the expansion (\ref{pser11}) with the general formula 
(\ref{5.11}), we find the decomposition of the AES over the 
orthonormal basis:
\begin{equation}
|k,\lambda,\vec{\beta}\rangle = {\cal N}^{-1/2} \sum_{n=0}^{\infty}
\left[ \frac{ n! \Gamma(2k) }{ \Gamma(2k+n) } \right]^{1/2}
P_{n}^{(-k+r-n,-k-r-n)}(x)\, \kappa^{n} |k,n\rangle \ .
\label{decomp11}
\end{equation}

The normalization factor is given by
\begin{equation}
{\cal N} = \sum_{n=0}^{\infty} \frac{ n! \Gamma(2k) }{ \Gamma(2k+n) }
\left| P_{n}^{(-k+r-n,-k-r-n)}(x) \right|^2 t^{n} \ , 
\label{Nseries11}
\end{equation}
where $t = |\kappa|^2$. The summation theorem for the Jacobi 
polynomials (Srivastava and Manocha, 1984, Sec.\ 2.3, Eqs.\ 60, 62) 
can be written in the form
\begin{equation}
\sum_{n=0}^{\infty} \frac{ n! \Gamma(\mu+\nu) 
}{ \Gamma(\mu+\nu+n) } \left| P_{n}^{(-\mu-n,-\nu-n)}(x)
\right|^2 t^{n} = S_{+}^{-\mu} S_{-}^{-\nu} 
F\left(\nu,\mu;\mu+\nu;-\frac{t}{S_{+}S_{-}} \right) \ ,   
\label{jp-sum11}
\end{equation}
for $\mu+\nu > 0$. Here, we have defined
\begin{equation}
S_{\pm} = 1 - |x \pm 1|^{2} t/4 = 1 - |\tau_{\mp}|^{2} \ .
\label{Sk-def}
\end{equation}
Therefore, we obtain the closed expression for the normalization 
factor:
\begin{equation}
{\cal N} = S_{+}^{-k+r} S_{-}^{-k-r}
F\left(k+r,k-r;2k;-\frac{t}{S_{+}S_{-}} \right) \ .
\end{equation}
If $|\tau_{-}| \geq 1$ or $|\tau_{+}| \geq 1$ [i.e., $r=k+l$ or 
$r=-(k+l)$, respectively], we can use the relation between the 
hypergeometric function and the Jacobi polynomials (Erd\'{e}lyi 
{\em et al.}, 1953, Vol.\ 2, Sec.\ 10.8). For $l$ being a nonnegative 
integer and $\alpha > -1$, this relation can be written in the form
\begin{equation}
F(-l,l+\alpha+\beta+1;\alpha+1;z) = \frac{ l! \Gamma(\alpha+1) 
}{ \Gamma(l+\alpha+1) } P_{l}^{(\alpha,\beta)}(1-2z) \ .
\label{hf-jpol-rel11}
\end{equation}
Then we obtain
\begin{equation}
{\cal N} = \frac{ l! \Gamma(2k) }{ \Gamma(2k+l) } 
S_{i}^{l} S_{i'}^{-2k-l} P_{l}^{(2k-1,0)}
\left( 1+\frac{2t}{S_{+}S_{-}} \right) \ ,
\end{equation}
where $(i,i') = (+,-)$ for $r=k+l$ and $(i,i') = (-,+)$ for 
$r=-(k+l)$. It can be easily verified that for $r = \pm k$
(i.e., $l = 0$), these formulas reduce to the corresponding results 
for the standard coherent state $|k,\zeta_{0}\rangle$ with 
$\zeta_{0} = -\tau_{\pm}$, respectively.

Analogously to the SU(2) case, the above analytic expressions can be 
used for calculations of quantum statistical properties of the SU(1,1) 
AES. Here, we derive analytic expressions for moments of the generator 
$K_{3}$. By using the property $K_{3} |k,n\rangle = (k+n) |k,n\rangle$ 
and formula (\ref{Nseries11}), moments of $K_{3}$ over the AES can be
expressed as derivatives of ${\cal N}$ with respect to $t$:
\begin{equation}
\langle K_{3} \rangle = \frac{t}{{\cal N}} 
\frac{\partial {\cal N}}{\partial t} + k \ , 
\end{equation}
\begin{equation}
(\Delta K_{3})^{2} = \frac{t^{2}}{{\cal N}} 
\frac{\partial^{2} {\cal N}}{\partial t^{2}} +
\frac{t}{{\cal N}} \frac{\partial {\cal N}}{\partial t}
- \left( \frac{t}{{\cal N}} \frac{\partial {\cal N}}{\partial t}
\right)^{2} \ . 
\end{equation}
By using the formula (Erd\'{e}lyi {\em et al.}, 1953, Vol.\ 1, 
Sec.\ 2.1.2)
\begin{equation}
\frac{ d F(a,b;c;z) }{ d z } = \frac{ab}{c} F(a+1,b+1;c+1;z)
\end{equation}
and the hypergeometric equation, we obtain exact analytic expressions 
for the moments of $K_{3}$:
\begin{equation}
\langle K_{3} \rangle = \frac{-k Y + r(S_{+}-S_{-})}{ S_{+}S_{-} }
+ \frac{ (k^{2}-r^{2}) Y t }{ 2 k S_{+}^{2} S_{-}^{2} }\, \Theta \ ,
\end{equation}
\begin{eqnarray}
(\Delta K_{3})^{2} & = & (k+r) \frac{ 1-S_{-} }{ S_{-}^{2} }
+ (k-r) \frac{ 1-S_{+} }{ S_{+}^{2} } - \frac{ (k^{2}-r^{2})
Y^{2} t }{ ( S_{+}S_{-} + t) S_{+}^{2} S_{-}^{2} } \nonumber \\
& & - \frac{ (k^{2}-r^{2})\, t }{ 2 k S_{+}^{3} S_{-}^{3} }
\left( \frac{ S_{+} S_{-} Y^{2} }{ S_{+}S_{-} + t } - 2 k Y^2 
+ Z \right) \Theta  - \frac{ (k^{2}-r^{2})^{2} Y^{2} t^{2} }{ 
4 k^{2} S_{+}^{4} S_{-}^{4} }\, \Theta^{2} \ .
\end{eqnarray}
Here, $Y$ and $Z$ are given by equations (\ref{Ydef}) and 
(\ref{Zdef}), respectively [but with $S_{\pm}$ of equation
(\ref{Sk-def})], and $\Theta$ is defined as
\begin{equation}
\Theta = \left[ F\left(k+r,k-r;2k;-\frac{t}{S_{+}S_{-}} \right) 
\right]^{-1} F\left(k+r+1,k-r+1;2k+1;-\frac{t}{S_{+}S_{-}} 
\right) \ .
\end{equation}
Note that the transition between the SU(2) and SU(1,1) cases can
be formally performed by the interchange:
\begin{equation}
j \leftrightarrow -k \ , \;\;\;\;\;\; 
m_{0} \leftrightarrow r \ , \;\;\;\;\;\;
t \leftrightarrow -t \ , \;\;\;\;\;\;
(j+|m_{0}|)\, \Omega \leftrightarrow 
\frac{k^{2}-r^{2}}{2k}\, \Theta \ .
\end{equation}
For $r = \pm (k+l)$ with $l > 0$, by using relation 
(\ref{hf-jpol-rel11}), we find 
\begin{equation}
\Theta = \frac{2k}{l} \left[ P_{l}^{(2k-1,0)} 
\left(1+\frac{2t}{S_{+}S_{-}} \right) \right]^{-1} 
P_{l-1}^{(2k,1)} \left(1+\frac{2t}{S_{+}S_{-}} \right) \ .
\end{equation}
For $r = \pm k$ (i.e., $l = 0$), we obtain $\Theta = 0$, and then 
we recover the known results for the SU(1,1) CS (Wodkiewicz and 
Eberly, 1985). The expressions for $\langle K_{3} \rangle$ and 
$(\Delta K_{3})^{2}$ are significantly simplified in the case 
$Y = 0$. This condition is satisfied if equation (\ref{Y0cond})
holds, e.g., for $\beta_{1} = a \beta_{2}$ where $a$ is any real 
number (including the case when $\beta_{1}$ or $\beta_{2}$
vanishes). Then we obtain
\begin{equation}
\langle K_{3} \rangle = \frac{h+1}{h-1}\, r \ , \label{k3ms}
\end{equation}
\begin{equation}
(\Delta J_{3})^{2} = \frac{ 2 k h }{ (h-1)^{2} } + 
\frac{ (k^{2}-r^{2}) h^2 t }{ k (h-1)^{4} }\, \Theta \ , \label{k3vs}
\end{equation}
where $h$ is defined by equation (\ref{hdef}). Note that the curve
$|\tau_{-}| = 1/|\tau_{+}|$ (that is, $Y=0$) lies in the allowed 
region of the parameter space (more specifically, in the
discrete-spectrum region), except for the forbidden point 
$|\tau_{-}| = |\tau_{+}| = 1$.

\subsection{Some special cases}

For $\beta_{+} = 0$ and $\beta_{3} \neq 0$, we obtain 
$\tau_{-} \rightarrow \infty$, so we cannot use formula
(\ref{5.20}). In this case the solution of equation (\ref{5.18}) is
\begin{equation}
\Lambda(k,\lambda,\vec{\beta};\zeta) = {\cal N}^{-1/2} 
\zeta^{l} (1 + \tau_{+} \zeta )^{-2k-l} \ , \label{kspec}
\end{equation}
where $\tau_{+} = \beta_{1}/\beta_{3}$ and $l = -k+\lambda/\beta_{3}$.
The condition of the analyticity requires $l = 0,1,2,\ldots$
[i.e., the spectrum $\lambda = (k+l)\beta_{3}$ is discrete] and 
$|\tau_{+}| < 1$. The decomposition over the orthonormal basis
and quantum statistical properties of the AES can be obtained in
this case by expanding the function (\ref{kspec}) into the power 
series. We find
\begin{equation}
|k,\lambda,\vec{\beta}\rangle = {\cal N}^{-1/2} \sum_{n=l}^{\infty} 
\left[ \frac{ n! \Gamma(2k+n) }{ l! \Gamma(2k+l) } \right]^{1/2} 
\frac{ (-\tau_{+})^{n-l} }{ (n-l)! } |k,n\rangle \ ,
\end{equation}
\begin{equation}
{\cal N} = F\left( l+1,l+2k;1;|\tau_{+}|^{2} \right)
= (1-|\tau_{+}|^{2})^{-2k-l} \, 
P_{l}^{(0,2k-1)} \left( \frac{ 1 + |\tau_{+}|^{2} }{
1-|\tau_{+}|^{2} } \right) \ .
\end{equation}
For $l=0$, the function (\ref{kspec}) represents the standard 
coherent state $|k,\zeta_{0}\rangle$ with $\zeta_{0} = -\tau_{+}$. 
The corresponding eigenvalue equation is 
\begin{equation}
( K_{3} - \zeta_{0} K_{+} ) |k,\zeta_{0}\rangle = 
k |k,\zeta_{0}\rangle \ .
\end{equation}

For $\beta_{-} = 0$ and $\beta_{3} \neq 0$, we can use the general 
results of sections \ref{su11general} and \ref{su11expansion}, with
$b = \beta_{3}$, $\tau_{+} = 0$ and $\tau_{-} = \beta_{3}/\beta_{1}$.
This gives $\kappa = -\tau_{-} = -\beta_{3}/\beta_{1}$, $x=1$, 
$S_{+} = 1-t$ (where $t=|\kappa|^{2}$) and $S_{-} = 1$. The 
corresponding analytic function is
\begin{equation}
\Lambda(k,\lambda,\vec{\beta};\zeta) = {\cal N}^{-1/2} 
( 1 + \tau_{-} \zeta )^{-k+r} \ ,   \label{5.35}   
\end{equation}
where $r=\lambda/\beta_{3}$. For $|\tau_{-}| < 1$, this function 
is always analytic and $\lambda$ can take any complex value. In the 
special case $r=-k$ (i.e., $\lambda = -\beta_{3} k$), the function
(\ref{5.35}) represents the standard coherent state 
$|k,\zeta_{0}\rangle$ with $\zeta_{0} = -\tau_{-}$. The corresponding
eigenvalue equation is
\begin{equation}
( K_{3} - \zeta_{0}^{-1} K_{-} ) |k,\zeta_{0}\rangle = 
-k |k,\zeta_{0}\rangle \ .
\end{equation}
For $|\tau_{-}| \geq 1$, the analyticity condition requires $r = k+l$ 
[i.e., the spectrum $\lambda = (k+l)\beta_{3}$ is discrete].
For $r = k$ ($l = 0$), we have $\Lambda(\zeta) = 1$ that represents 
the state $|k,0\rangle$.

We next consider the degenerate case $b=0$. If $\beta_{+}$ vanishes, 
then $\beta_{3}$ vanishes too, and the corresponding algebra 
element is just $K_{+}$. It can be easily verified that this 
operator has not any eigenstate. If $\beta_{-}$ vanishes, 
then $\beta_{3}$ vanishes too, and the corresponding algebra 
element is just $K_{-}$. Its eigenstates are represented by
the analytic function
\begin{equation}
\Lambda(k,\lambda,\vec{\beta};\zeta) = {\cal N}^{-1/2} 
\exp( \lambda \zeta ) \ .
\end{equation}
The eigenvalue $\lambda$ can take any complex value. By using 
equation (\ref{5.13}), we find the normalization factor
\begin{equation}
{\cal N} = \Gamma(2k) I_{2k-1}(2|\lambda|) \lambda^{1-2k}\ ,   
\label{5.41}
\end{equation}
where $I_{\nu}(x)$ is the $\nu$-order modified Bessel function of 
the first kind. The eigenstates of the lowering operator $K_{-}$ 
were first constructed by Barut and Girardello (1971). 
The analytic representation based on these states will be discussed 
in section \ref{sub:B}. 

If $b=0$ and $\beta_{3} \neq 0$, the solution of equation 
(\ref{5.18}) is 
\begin{equation}
\Lambda(k,\lambda,\vec{\beta};\zeta) = {\cal N}^{-1/2} 
( 1 + \tau \zeta )^{-2k} \exp\left( - \frac{ 2\lambda }{ \beta_{3} }
\frac{ 1 }{ 1 + \tau \zeta } \right)   \ ,   \label{la-degen}
\end{equation}
where 
\begin{equation}
\tau = \frac{ 2\beta_{-} }{ \beta_{3} } = 
\frac{ \beta_{3} }{ 2\beta_{+} }  \ .   \label{tau-def}
\end{equation}
The condition of the analyticity requires $|\tau| < 1$. If this
condition is satisfied, $\lambda$ can take any complex value.
The decomposition of the corresponding AES over the orthonormal
basis is obtained in section \ref{sub:B} by using the 
Barut-Girardello analytic representation [see equations (\ref{dexp})
and (\ref{dnorm})]. Analogously to the general case, there are
operators whose eigenstates are unnormalizable in sense of equation
(\ref{5.13}), but these states can be orthonormalized by the delta
function. An example is the operator $K_{1} + K_{3}$, for which
$|\tau| = 1$ (Lindblad and Nagel, 1970; Nagel, 1995).
In a special case $\lambda = 0$, the analytic function is
\begin{equation}
\Lambda(k,\lambda=0,\vec{\beta};\zeta) = {\cal N}^{-1/2} 
( 1 + \tau \zeta )^{-2k} \ .
\label{5.32}   
\end{equation}
This function represents the standard coherent state 
$|k,\zeta_{0}\rangle$ with $\zeta_{0} = -\tau$
and ${\cal N} = (1-|\zeta_{0}|^{2})^{-2k}$. For example, 
we can choose $\vec{\beta} = \Bigl(1+\zeta_{0}^{2}, 
-i(1-\zeta_{0}^{2}), -2\zeta_{0} \Bigr)$. Then the standard CS 
satisfy the eigenvalue equation
\begin{equation}
( K_{-} - 2 \zeta_{0} K_{3} + \zeta_{0}^{2} K_{+} ) 
|k,\zeta_{0}\rangle = 0 \ .   \label{5.34}
\end{equation}

\subsection{The SU(1,1) intelligent states}

The SU(1,1) generalized IS were defined by Trifonov (1994) as the 
eigenstates of the operator $\eta K_{1} - i K_{2}$ [see equation 
(\ref{2.18})]. In our notation, the generalized IS are the AES with 
$\vec{\beta} = (\eta,-i,0)$. For $\eta^{2} \neq 1$,
the corresponding analytic function is given by the particular case 
of equation (\ref{5.20}) with $b = \sqrt{1-\eta^{2}}$ and
\begin{equation}
\tau_{\pm} = \mp \sqrt{\frac{ 1-\eta }{ 1+\eta }}  \ . 
\label{5.42}   \end{equation}
The condition of the analyticity requires 
\begin{equation}
\left| \frac{ 1-\eta }{ 1+\eta } \right| < 1 \;
\Leftrightarrow \; {\rm Re}\, \eta > 0 \ .  \label{5.43}
\end{equation}
In the allowed region of the parameter space (${\rm Re}\, \eta > 0$), 
all the results of section \ref{su11expansion} are valid, with
$\kappa = 2\tau_{+}$, $x = 0$, and $S_{+} = S_{-} = 1-|\tau_{+}|^{2}$.
For $\eta = 0$, the algebra element is $K_{2}$ whose eigenstates do 
not possess a finite norm.
The generalized intelligent state is also the standard coherent state
when $r = \pm k$. The corresponding coherent-state amplitude is
$\zeta_{0} = -\tau_{\pm}$, respectively. Since $\eta$ is a complex 
number, $\zeta_{0}$  can acquire any complex value in the unit disk 
(excluding zero because $\eta \neq 1$). For $\eta = 1$, the generalized 
IS are reduced to the BG states (the eigenstates of $K_{-}$). When the 
eigenvalue vanishes, $\lambda = 0$, the BG states degenerate to the 
state $|k,0\rangle$ that is the standard coherent state with 
$\zeta_{0} = 0$. Therefore, the standard set of the SU(1,1) CS is a 
subset of the SU(1,1) generalized IS.

We also briefly discuss three types of the SU(1,1) ordinary IS. 
The $K_{1}$-$K_{2}$ IS are determined, according to equation 
(\ref{2.19}), as the eigenstates of the operator 
$K_{1} +i\gamma K_{2}$, where $\gamma$ is a real parameter. 
These states are the SU(1,1) AES with $\vec{\beta} = (1,i\gamma,0)$. 
For $\gamma=1$, the solution does not exist since the raising operator 
$K_{+}$ has not any eigenstate. For $\gamma=-1$, we obtain the 
BG states. For $\gamma^{2} \neq 1$, the corresponding analytic 
function is given by the particular case of equation (\ref{5.20})
with $b = \sqrt{\gamma^{2}-1}$ and
\begin{equation}
\tau_{\pm} = \pm \sqrt{ \frac{ \gamma+1 }{ \gamma-1 } } \ .  
\label{5.46}   
\end{equation}
The condition of the analyticity requires 
\begin{equation}
\left| \frac{ \gamma+1 }{ \gamma-1 } \right| < 1 \;
\Leftrightarrow \; \gamma < 0 \ .   \label{5.47}
\end{equation}
All the results of section \ref{su11expansion} are valid here with
$\kappa = 2\tau_{+}$ and $x = 0$. We also find that $S_{+} = S_{-} = 
2|\gamma|/(1+|\gamma|)$ for $|\gamma|<1$, and $S_{+} = S_{-} = 
2/(1+|\gamma|)$ for $|\gamma|>1$. For $\gamma = 0$, the algebra 
element is $K_{1}$ whose eigenstates do not possess a finite norm.
The $K_{1}$-$K_{2}$ IS coincide with the standard CS for 
$r = \pm k$. The corresponding coherent-state amplitude is 
$\zeta_{0} = -\tau_{\pm}$, respectively. Since $\gamma$ is real, 
$\zeta_{0}$ is real for $|\gamma| > 1$ and pure imaginary for 
$|\gamma| < 1$. Therefore the set of the $K_{1}$-$K_{2}$ IS and 
the standard set of the SU(1,1) CS have an intersection. 

The $K_{1}$-$K_{3}$ IS are determined, according to equation 
(\ref{2.19}), as the eigenstates of the operator 
$K_{1} +i\gamma K_{3}$. These states are the SU(1,1) AES with 
$\vec{\beta} = (1,0,i\gamma)$. The corresponding analytic function 
is given by the particular case of equation (\ref{5.20})
with $b = i \sqrt{\gamma^{2}+1}$ and
\begin{equation}
\tau_{\pm} = - i \left(\gamma \pm \sqrt{\gamma^{2}+1} \right)^{-1} \ .      
\label{5.46p}   
\end{equation}
The analyticity condition requires $r = k+l$ for $|\tau_{-}| > 1$
(i.e., for $\gamma > 0$), and $r = -(k+l)$ for $|\tau_{+}| > 1$
(i.e., for $\gamma < 0$). Here, as usual, $l = 0,1,2,\ldots$.
This condition can be expressed in the form
\begin{equation}
\lambda = i ({\rm sgn}\, \gamma) (k+l) \sqrt{\gamma^{2}+1} \ ,
\label{areq}
\end{equation}
(For $\gamma = 0$, we have $|\tau_{+}| = |\tau_{-}| = 1$, and the
algebra element is $K_{1}$ with unnormalizable eigenstates.)
All the results of section \ref{su11expansion} are valid here with
$\kappa = -2 i \sqrt{\gamma^{2}+1}$ and 
$x = \gamma/\sqrt{\gamma^{2}+1}$. Note that $|\tau_{+}\tau_{-}| = 1$,
and therefore we can use simple expressions (\ref{k3ms}) and 
(\ref{k3vs}), with $h = 2 \gamma^{2} + 2 \sqrt{\gamma^{2}+1} + 1$.
The intersection between the $K_{1}$-$K_{3}$ IS and the standard CS 
is obtained for $l=0$, i.e., $\lambda = \pm i k \sqrt{\gamma^{2}+1}$. 
The corresponding coherent-state amplitude is 
$\zeta_{0} = -\tau_{\pm}$, respectively. 
Since $\gamma$ is real, $\zeta_{0}$ is pure imaginary.

The $K_{2}$-$K_{3}$ IS are determined, according to 
equation (\ref{2.19}), as the eigenstates of the operator 
$K_{2} +i\gamma K_{3}$. These states are the SU(1,1) AES with 
$\vec{\beta} = (0,1,i\gamma)$. The corresponding analytic function 
is given by the particular case of equation (\ref{5.20})
with $b = i \sqrt{\gamma^{2}+1}$ and
\begin{equation}
\tau_{\pm} = - \left(\gamma \pm \sqrt{\gamma^{2}+1} \right)^{-1}  \ .  
\label{5.46pp}   
\end{equation}
The analyticity condition here is the same as in the preceding case
[see equation (\ref{areq})]. 
All the results of section \ref{su11expansion} are valid here with
$\kappa = -2\sqrt{\gamma^{2}+1}$ and $x = \gamma/\sqrt{\gamma^{2}+1}$.
Once again, $|\tau_{+}\tau_{-}| = 1$, and therefore we can use simple 
expressions (\ref{k3ms}) and (\ref{k3vs}), with 
$h = 2 \gamma^{2} + 2 \sqrt{\gamma^{2}+1} + 1$.
The intersection between the $K_{2}$-$K_{3}$ IS and the standard CS 
is obtained for $l=0$, i.e., $\lambda = \pm i k \sqrt{\gamma^{2}+1}$. 
The corresponding coherent-state amplitude is
$\zeta_{0} = -\tau_{\pm}$, respectively. 
Since $\gamma$ is real, $\zeta_{0}$ is also real.

\subsection{The Barut-Girardello analytic representation}
\label{sub:B}

The BG states (Barut and Girardello, 1971) are defined as the 
eigenstates of the lowering operator $K_{-}$:
\begin{equation}
K_{-} |k,z\rangle = z |k,z\rangle \ , 
\end{equation}
where $z$ is an arbitrary complex number. The expansion of these states
over the orthonormal basis is
	\begin{equation}
|k,z\rangle = \frac{z^{k-1/2}}{\sqrt{I_{2k-1}(2|z|)}} 
\sum_{n = 0}^{\infty} \frac{z^{n}}{\sqrt{n!\Gamma(2k+n)}} 
|k,n\rangle \ , \label{5.58}
	\end{equation}
where the eigenvalue $z$ is an arbitrary complex number. The BG 
states are normalized but they are not orthogonal to each other:
	\begin{equation}
\langle k,z_{1}|k,z_{2} \rangle = \frac{ I_{2k-1} 
(2\sqrt{z_{1}^{\ast}z_{2}}) 
}{ \sqrt{ I_{2k-1}(2|z_{1}|) I_{2k-1}(2|z_{2}|) } } \ .  \label{5.59}
	\end{equation}
The identity resolution is
	\begin{equation}
\int d\mu(k,z) |k,z \rangle\langle k,z| = I \ , \mbox{\hspace{1.0cm}} 
d\mu(k,z) = \frac{2}{\pi} K_{2k-1}(2|z|) 
I_{2k-1}(2|z|) d^{2}\! z \ ,  \label{5.60}
	\end{equation}
where $K_{\nu}(x)$ is the $\nu$-order modified Bessel function of 
the second kind. Thus the BG states form an overcomplete set. 
Therefore, for any state $| \Psi \rangle = \sum_{n=0}^{\infty} 
c_{n} |k,n\rangle$ in the Hilbert space, 
one can construct the analytic function
	\begin{equation}
f(z) =  \frac{\sqrt{I_{2k-1}(2|z|)}}{z^{k-1/2}} \langle 
k,z^{\ast}|\Psi\rangle = \sum_{n = 0}^{\infty} 
\frac{c_{n} }{ \sqrt{n!\Gamma(2k+n)} } z^{n} \ .  \label{5.61}
	\end{equation}
Then the state $| \Psi \rangle$ can be represented in the BG basis
\begin{equation}
| \Psi \rangle = \int d\mu(k,z) \frac{(z^{\ast})^{k-1/2}}{
\sqrt{I_{2k-1}(2|z|)}} f(z^{\ast}) |k,z \rangle \ ,  \label{5.62}
\end{equation}
\begin{equation}
\langle\Psi|\Psi\rangle = \int d\mu(k,z) \frac{|z|^{2k-1} }{ 
I_{2k-1}(2|z|) } |f(z^{\ast})|^{2} < \infty \ .  \label{5.63}
\end{equation}
The standard coherent state $|k,\zeta\rangle$ is represented by the 
function 
\begin{equation}
{\cal F}(k,\zeta;z) = \frac{\sqrt{I_{2k-1}(2|z|)}}{z^{k-1/2}} 
\langle k,z^{\ast}|k,\zeta\rangle =
\frac{ (1-|\zeta|^{2})^{k} }{ \sqrt{\Gamma(2k)} } \exp(\zeta z) \ .   
\label{5.64}
\end{equation}
As has been recently shown (Brif {\em et al.}, 1996), the BG
representation and the analytic representation in the unit disk
are related through a Laplace transform.

The operators ${K}_{\pm}$ and ${K}_{3}$ act in the Hilbert space of 
analytic functions $f(z)$ as linear  differential operators
	\begin{equation}
{K}_{+} =  z \ , \mbox{\hspace{0.8cm}} {K}_{-} = 2k \frac{d}{dz} + z 
\frac{d^{2}}{dz^{2}} \ , \mbox{\hspace{0.8cm}}
{K}_{3} = z \frac{d}{dz} + k  \ .   \label{5.65}
	\end{equation}
By introducing the analytic function
\begin{equation}
\Lambda(k,\lambda,\vec{\beta};z) = \frac{\sqrt{I_{2k-1}(2|z|)}}{
z^{k-1/2}} \langle k,z^{\ast}|k,\lambda,\vec{\beta}\rangle \ ,
\label{5.66}    \end{equation}
the eigenvalue equation (\ref{5.16}) for the SU(1,1) AES 
$|k,\lambda,\vec{\beta}\rangle$ can be converted into the
second-order linear homogeneous differential equation
\begin{equation}
[\beta_{+} z] \frac{d^{2}}{dz^{2}} \Lambda(k,\lambda,\vec{\beta};z) 
+ [\beta_{3} z + 2k \beta_{+}] \frac{d}{dz}
\Lambda(k,\lambda,\vec{\beta};z) + [\beta_{-} z + k \beta_{3} 
-\lambda] \Lambda(k,\lambda,\vec{\beta};z) = 0 \ .  \label{5.67}
\end{equation}
This equation can be transformed into the Kummer equation for the 
confluent hypergeometric function $\Phi(a;c;x)$ or into the 
Bessel equation, depending on the values of the parameters. 
Using the general results of Erd\'{e}lyi {\em et al.} 
(1953, Vol.\ 1, Sec.\ 6.2), we find the solution of equation 
(\ref{5.67}). 

We first consider the general case $b \neq 0$. 
For $\beta_{+} \neq 0$, we have two independent solutions:
\begin{eqnarray}
& & \Lambda(k,\lambda,\vec{\beta};z) = \exp( -\tau_{\pm} z ) \,
\Phi \Bigl( k \mp r ; 2k ; \mp b z / \beta_{+}  \Bigr) \ ,  
\label{5.68} \\
& & \tilde{\Lambda}(k,\lambda,\vec{\beta};z) = 
\exp( -\tau_{\pm} z )\, z^{1-2k} \, \Phi \Bigl( \tilde{k} \mp r ; 
2\tilde{k} ; \mp b z / \beta_{+}  \Bigr) \ ,  
\label{5.68p}
\end{eqnarray}
where $\tau_{\pm}$ is defined by equation (\ref{taupm-def}), 
$r = \lambda/b$, and $\tilde{k} = 1-k$. The first solution $\Lambda$ 
is always analytic, but the second solution $\tilde{\Lambda}$ is not 
(except for $k=\frac{1}{2}$ when the two solutions coincide).
[Another special case is $k=\frac{1}{4}$ and $\tilde{k} =\frac{3}{4}$
(or vice versa). These representations do not belong to the discrete 
series, but they occur for the so-called two-photon realization
of the su(1,1) Lie algebra. This realization is important in quantum 
optics, as it provides the mathematical model of squeezing of the 
single-mode quantized light field. The corresponding analytic
representations have been recently studied in detail (Brif {\em et 
al.}, 1996; Brif, 1996).] Here, we consider only the discrete-series
representations ($k$ is a positive integer or half-integer) and hence 
use only the first solution $\Lambda(z)$. Also, the solution should 
be normalizable, i.e., the integral in equation (\ref{5.63}) 
must be convergent. The convergence of this integral can be easily
analyzed by using the asymptotic expansion of the integrand.
It is remarkable that the normalization condition for the function
$\Lambda(z)$ of equation (\ref{5.68}) is equivalent to the
analyticity condition for the function $\Lambda(\zeta)$ of equation
(\ref{5.20}) [see the discussion after equation (\ref{r-def})
and Fig.~1]. Actually, these functions are related through a Laplace 
transform (Brif {\em et al.}, 1996). 
The upper and lower signs in equation (\ref{5.68}) are equivalent, 
because the confluent hypergeometric function can be written in two 
equivalent forms which are related by Kummer's transformation 
(Erd\'{e}lyi {\em et al.}, 1953, Vol.\ 1, Sec.\ 6.3):
\begin{equation}
\Phi(a;c;x) = e^{x} \Phi(c-a;c;-x) \ .
\end{equation}
Kummer's transformation in equation (\ref{5.68}) is equivalent to
the replacement $b \leftrightarrow -b$.
For $\lambda = \pm k b$, the AES coincide with the standard 
CS. Then equation (\ref{5.68}) reads
\begin{equation}
\Lambda(k,\lambda,\vec{\beta};z) = {\cal N}^{-1/2} \exp\left(  
- \tau_{\pm} z \right) \ .
\label{5.70}   \end{equation} 
The corresponding coherent-state amplitude is 
$\zeta_{0} = - \tau_{\pm}$.

In the case $\beta_{+} = 0$ and $\beta_{3} \neq 0$, the solution of 
equation (\ref{5.67}) is
\begin{equation}
\Lambda(k,\lambda,\vec{\beta};z) = {\cal N}^{-1/2} z^{l} 
\exp( -\tau_{+} z ) \ ,   \label{5.76}
\end{equation}
where $\tau_{+} = \beta_{1} / \beta_{3}$ and 
$l = -k + \lambda/\beta_{3}$.
The condition of the analyticity requires $l = 0,1,2,\ldots$,
and the normalization condition is $|\tau_{+}| < 1$. This is
in full agreement with the analyticity condition for the
function $\Lambda(\zeta)$ of equation (\ref{kspec}).

We next consider the degenerate case $b=0$. If $\beta_{+}$ vanishes, 
equation (\ref{5.67}) has not any nontrivial analytic solution, and
consequently the operator $K_{+}$ has not any eigenstate. For
$\beta_{+} \neq 0$, the analytic solution of equation (\ref{5.67}) is
\begin{equation}
\Lambda(k,\lambda,\vec{\beta};z) = {\cal N}^{-1/2} 
\left( \lambda' z \right)^{1/2 - k} 
I_{2k-1} \left( 2\sqrt{ \lambda' z } \right) 
\exp( - \tau z ) \ ,  \label{5.72}
\end{equation}
where $\tau$ is defined by equation (\ref{tau-def}) 
and $\lambda' = \lambda/\beta_{+}$.
The normalization condition requires $|\tau| < 1$, in accordance
with the analyticity condition for the function $\Lambda(\zeta)$ of
equation (\ref{la-degen}). The decomposition of the AES with $b=0$
over the orthonormal basis and the corresponding normalization factor 
can be obtained by expanding the analytic function $\Lambda(z)$ of 
equation (\ref{5.72}) into the power series. By using generating
functions for the Laguerre polynomials $L_{n}^{\alpha}(x)$
(Erd\'{e}lyi {\em et al.}, 1953, Vol.\ 2, Sec.\ 10.12;
Srivastava and Manocha, 1984, Sec.\ 2.5), we find
\begin{equation}
|k,\lambda,\vec{\beta}\rangle = {\cal N}^{-1/2} \sum_{n=0}^{\infty}
\left[ \frac{n!}{\Gamma(2k+n)} \right]^{1/2} L_{n}^{2k-1}
(\lambda' / \tau)\, (-\tau)^{n}\, |k,n\rangle \ ,  \label{dexp}
\end{equation}
\begin{equation}
{\cal N} = \frac{ |\lambda'|^{1-2k} }{1 - |\tau|^{2} }\, 
I_{2k-1}\left( \frac{ 2 |\lambda'| }{1 - |\tau|^{2} } \right) 
\exp\left[ -\frac{ 2|\tau|^{2}\, {\rm Re}\, (\lambda' / \tau) 
}{ 1-|\tau|^{2} } \right] \ .  \label{dnorm}
\end{equation}
In a special case $\lambda = 0$,
the corresponding AES are the standard CS $|k,\zeta_{0}\rangle$ 
with $\zeta_{0} = - \tau$. If $\beta_{-}$ vanishes, $\beta_{3}$ 
vanishes too, and the corresponding algebra element is just
$K_{-}$. Then equation (\ref{5.72}) reads
\begin{equation}
\Lambda(k,\lambda,\vec{\beta};z) = {\cal N}^{-1/2} \frac{ 
I_{2k-1}(2\sqrt{\lambda z}) }{ (\lambda z)^{k-1/2} } \ .  \label{5.74}
\end{equation}
This function represents the BG state $|k,z_{0}\rangle$ with 
$z_{0} = \lambda$, and the normalization factor is
\begin{equation}
{\cal N} = z_{0}^{1-2k}\, I_{2k-1}(2|z_{0}|)  \ .   \label{5.75}
\end{equation}

One can also use the general solution (\ref{5.68}) in order to 
consider the particular case of the SU(1,1) IS. The BG 
representation of the SU(1,1) generalized IS was first obtained by 
Trifonov (1994). His results can be reproduced by taking 
$\vec{\beta} = (\eta,-i,0)$ that corresponds to the algebra element 
$\eta K_{1} - iK_{2}$. It is not difficult to obtain also the BG 
analytic representation for different types of the SU(1,1) ordinary
IS.

\section{\uppercase{Conclusions}}
\label{sec:Conc}

In this paper we have shown that the algebra-eigenstate formalism
unifies the descriptions of various kinds of states within a common
frame. This clarifies relations between different types of states
and the physical basis of their mathematical properties. The use
of an analytic representation enables us to write a linear 
homogeneous differential equation that determines all the kinds
of the AES. Often, this is a first-order equation that can be 
immediately integrated, and then we derive analytic functions 
representing the AES. These functions yield all the information 
about the AES. For example, we have presented here a method for the 
calculation of exact analytic expressions for quantum statistical 
properties of the AES. This method should be useful in physical
applications, especially in the field of quantum optics.
On the other hand, the analytic representations provide us with a 
simple and effective way to find different subsets of the AES and to 
analyse the relations between them. We have determined, for instance, 
the conditions to obtain the standard CS and the generalized and 
ordinary IS.

In the present work we have concentrated on the most elementary 
simple Lie groups, but the theory of the AES is in general 
applicable to arbitrary Lie groups describing a wide class of 
quantum systems. Therefore, the algebra-eigenstate formalism can
find applications in many fields of modern quantum physics.
We mention, for example, quantum optics where the AES can provide a 
general view on the problem of squeezing with practical applications 
to the reduction of quantum fluctuations and the improvement of the 
accuracy of interferometric measurements.

\section*{\uppercase{Acknowledgements}}

I wish to thank my colleagues A. Berengolts, K. Gokhberg,
A. Kenis and M. Lublinsky for their interest in my work and
many stimulating discussions. The financial help from the
Technion is gratefully acknowledged.

\newpage

\section*{\uppercase{References}}
\begin{description}
\item Agarwal, G. S. (1988). {\em Journal of Optical Society of 
America B}, {\bf 5}, 1940.
\item Agarwal, G. S., and Puri R. R. (1990). {\em Physical Review A}, 
{\bf 41}, 3782.
\item Aragone, C., Guerri, G., Salamo, S., and Tani, J. L. (1974).
{\em Journal of Physics A}, {\bf 7}, L149.
\item Aragone, C., Chalbaud, E., and Salamo, S. (1976). {\em Journal
of Mathematical Physics}, {\bf 17}, 1963.
\item Bargmann, V. (1947). {\em Annals of Mathematics}, {\bf 48}, 568.
\item Bargmann, V. (1961). {\em Communications in Pure and Applied
Mathematics}, {\bf 14}, 187.
\item Barut, A. O., and Girardello, L. (1971). {\em Communications in
Mathematical Physics}, {\bf 21}, 41.
\item Berezin, F. A. (1975). {\em Communications in Mathematical 
Physics}, {\bf 40}, 153.
\item Bergou, J. A., Hillery, M., and Yu, D. (1991). {\em Physical 
Review A}, {\bf 43}, 515.
\item Brif, C. (1995). {\em Quantum and Semiclassical Optics},
{\bf 7}, 803.
\item Brif, C. (1996). {\em Annals of Physics}, {\bf 251}, 180.
\item Brif, C., and Ben-Aryeh, Y. (1994a). {\em Journal of Physics A}, 
{\bf 27}, 8185.
\item Brif, C., and Ben-Aryeh, Y. (1994b). {\em Quantum Optics},
{\bf 6}, 391.
\item Brif, C., and Ben-Aryeh, Y. (1996). {\em Quantum and 
Semiclassical Optics}, {\bf 8}, 1.
\item Brif, C., and Mann, A. (1996a). {\em Physics Letters A}, 
{\bf 219}, 257.
\item Brif, C., and Mann, A. (1996b). {\em Physical Review A},
{\bf 54}, 4505.
\item Brif, C., Vourdas, A., and Mann, A. (1996). 
{\em Journal of Physics A}, {\bf 29}, 5873.
\item Bu\v{z}ek, V. (1990). {\em Journal of Modern Optics}, 
{\bf 37}, 303.
\item Dodonov, V. V., Malkin, I. A., and Man'ko, V. I. (1974).
{\em Physica}, {\bf 72}, 597.
\item Dodonov, V. V., Kurmyshev, E. V., and Man'ko, V. I. (1980).
{\em Physics Letters A}, {\bf 79}, 150.
\item Erd\'{e}lyi, A. {\em et al.}, eds. (1953). {\em Higher 
Transcendental Functions: Bateman Manuscript Project}, 
McGraw-Hill, New York.
\item Fock, V. A. (1928). {\em Zeitschrift f\"{u}r Physik}, 
{\bf 49}, 339.
\item Gerry, C. C., and Grobe, R. (1995). {\em Physical Review A}, 
{\bf 51}, 4123.
\item Gilmore, R. (1972). {\em Annals of Physics}, {\bf 74}, 391.
\item Gilmore, R. (1974). {\em Journal of Mathematical Physics}, 
{\bf 15}, 2090.
\item Glauber, R. J. (1963). {\em Physical Review}, {\bf 131}, 2766.
\item Hillery, M. (1987). {\em Physical Review A}, {\bf 36}, 3796.
\item Hillery, M. (1989). {\em Physical Review A}, {\bf 40}, 3147.
\item Hillery, M, and Mlodinow, L. (1993). {\em Physical Review A}, 
{\bf 48}, 1548.
\item Jackiw, R. (1968). {\em Journal of Mathematical Physics},
{\bf 9}, 339.
\item Klauder, J. R., and Skagerstam, B. S. (1985). {\em Coherent 
states: Applications in Physics and Mathematical Physics}, 
World Scientific, Singapore.
\item Lindblad, G., and Nagel, B. (1970). {\em Annales de 
l'Institut Henri Poincar\'{e} A}, {\bf 13}, 27.
\item Luis, A., and Pe\v{r}ina, J. (1996). 
{\em Physical Review A}, {\bf 53}, 1886.
\item Nagel, B. (1995). In {\em Modern Group Theoretical Methods
in Physics}, J. Bertrand {\em et al.}, eds., Kluwer, Dordrecht, 
pp. 211-220.
\item Nieto, M. M., and Truax, D. (1993). {\em Physical Review 
Letters}, {\bf 71}, 2843.
\item Perelomov, A. M. (1972). {\em Communications in Mathematical 
Physics}, {\bf 26}, 222.
\item Perelomov, A. M. (1977). {\em Soviet Physics Uspekhi}, 
{\bf 20}, 703.
\item Perelomov, A. M. (1986). {\em Generalized Coherent States and 
Their Applications}, Springer, Berlin.
\item Prakash, G. S., and Agarwal, G. S. (1994). 
{\em Physical Review A}, {\bf 50}, 4258.
\item Prakash, G. S., and Agarwal, G. S. (1995). 
{\em Physical Review A}, {\bf 52}, 2335.
\item Puri, R. R. (1994). {\em Physical Review A}, {\bf 49}, 2178.
\item Puri, R. R., and Agarwal, G. S. (1996). 
{\em Physical Review A}, {\bf 53}, 1786.
\item Rasetti, M. (1975). {\em International Journal of Theoretical
Physics}, {\bf 13}, 425.
\item Ruschin, S., and Ben-Aryeh, Y. (1976). {\em Physics Letters A},
{\bf 58}, 207.
\item Schr\"{o}dinger, E. (1926). {\em Die Naturwissenschaften}, 
{\bf 14}, 664. 
\item Segal, I. E. (1962). {\em Illinois Journal of Mathematics},
{\bf 6}, 500.
\item Solomon, A. I. (1971). {\em Journal of Mathematical Physics},
{\bf 12}, 390.
\item Srivastava, H. M., and Manocha, H. L. (1984). {\em A Treatise
on Generating Functions}, Ellis Horwood, Chichester.
\item Stoler, D. (1970). {\em Physical Review D}, {\bf 1}, 3217.
\item Stoler, D. (1971). {\em Physical Review D}, {\bf 4}, 2308.
\item Trifonov, D. A. (1994). {\em Journal of Mathematical Physics},
{\bf 35}, 2297.
\item Trifonov, D. A. (1996a). Preprint, quant-ph/9609001.
\item Trifonov, D. A. (1996b). Preprint, quant-ph/9609017.
\item Vanden-Bergh, G., and DeMeyer, H. (1978). {\em Journal of
Physics A}, {\bf 11}, 1569.
\item Vilenkin, N. Ya. (1968). {\em Special Functions and the 
Theory of Group Representations}, American Mathematical Society,
Providence, Rhode Island.
\item Weyl, H. (1950). {\em The Theory of Groups and Quantum 
Mechanics}, Dover, New York.
\item Wodkiewicz, K., and Eberly, J. H. (1985). {\em Journal of
Optical Society of America B}, {\bf 2}, 458.
\item Yu, D., and Hillery, M. (1994). {\em Quantum Optics}, 
{\bf 6}, 37.
\item Yuen, H. (1976). {\em Physical Review A}, {\bf 13}, 2226.
\item Zhang, W. M., Feng, D. H., and Gilmore, R. (1990). {\em Reviews
of Modern Physics}, {\bf 62}, 867.
\end{description}

\end{document}